\documentclass[conference]{IEEEtran}
\ifCLASSINFOpdf
\else
\fi
\usepackage{tikz}
\usepackage{multirow}
\usepackage{multicol}
\usepackage{amsmath}
\usepackage{amssymb}
\usepackage{nicefrac}
\usepackage{xspace}
\usepackage{caption}
\usepackage{subcaption}
\usepackage{placeins}

\usepackage{algorithm}
\usepackage{algpseudocode}
\usepackage{numprint}
\usepackage{graphicx}
\graphicspath{ {./images/} }
\usepackage{xcolor}
\usepackage{tcolorbox}
\usepackage{tabularx}
\usepackage{array}
\usepackage{colortbl}
\let\labelindent\relax
\usepackage{enumitem}
\tcbuselibrary{skins}
\usepackage{hyperref}
\newcolumntype{Y}{>{\raggedleft\arraybackslash}X}
\newcolumntype{Z}{>{\centering\arraybackslash}X}

\tcbuselibrary{raster}
\tcbset{tab1/.style={fonttitle=\bfseries\large,fontupper=\normalsize\sffamily,
colback=yellow!10!white,colframe=red!75!black,colbacktitle=green!40!white,
coltitle=black,center title,freelance,frame code={
\foreach \n in {north east,north west,south east,south west}
{\path [fill=red!75!black] (interior.\n) circle (3mm); };},}}

\tcbset{tab2/.style={enhanced,fonttitle=\bfseries,fontupper=\normalsize\sffamily,
colback=yellow!5!white,colframe=black!50!black,colbacktitle=gray!20!white,
coltitle=black,center title}}

\tcbset{tab3/.style={enhanced,fonttitle=\bfseries\small,fontupper=\normalsize\sffamily\footnotesize,
colback=yellow!5!white,colframe=black!50!black,colbacktitle=gray!20!white,
coltitle=black,center title}}
\tcbset{tab4/.style={enhanced,fonttitle=\bfseries\small,fontupper=\normalsize\sffamily\footnotesize,
colback=yellow!5!white,colframe=black!50!black,colbacktitle=gray!20!white,
coltitle=black,center title, width=1.2\columnwidth}}

\tcbset{tab5/.style={enhanced,fonttitle=\bfseries\small,fontupper=\normalsize\sffamily\footnotesize,
colback=yellow!5!white,colframe=black!50!black,colbacktitle=gray!20!white,
coltitle=black,center title, width=1.4175\columnwidth}}

\usepackage{makecell}

\usepackage{newfloat}
\DeclareFloatingEnvironment[fileext=frm,placement={!tb},name=Frame]{tablefloat}

\newcommand{\ournameNoSpace}{\mbox{CrowdGuard}}
\newcommand{\ournameGen}{\ournameNoSpace's\xspace}
\newcommand{\ourname}{\ournameNoSpace\xspace}
\newcommand{\paperTitle}{\ournameNoSpace: Federated Backdoor Detection in Federated Learning} 
\newcommand{\sect}{Sect.~}
\newcommand{\fig}{Fig.~}
\newcommand{\app}{App.~}
\newcommand{\alg}{Alg.~}
\newcommand{\equ}{Eq.~}
\newcommand{\tab}{Tab.~}
\newcommand{\fedavg}{\mbox{FedAVG}\xspace}
\newcommand{\etal}{\emph{et~al.}\xspace}
\newcommand{\nonIidNoSpace}{non-IID}
\newcommand{\nonIidBigNoSpace}{Non-IID}
\newcommand{\nonIidBig}{\nonIidBigNoSpace\xspace}
\newcommand{\iidNoSpace}{IID}
\newcommand{\iid}{\iidNoSpace\xspace}
\newcommand{\nonIid}{\nonIidNoSpace\xspace}
\newcommand{\clientMetricLong}{Hidden Layer Backdoor Inspection Metric\xspace}
\newcommand{\clientMetricNoSpace}{HLBIM}
\newcommand{\clientMetric}{\clientMetricNoSpace\xspace}
\newcommand{\serverNoSpace}{\ensuremath{\mathcal{S}}}
\newcommand{\server}{\serverNoSpace\xspace}
\newcommand{\adversaryNoSpace}{\ensuremath{\mathcal{A}}}
\newcommand{\adversary}{\adversaryNoSpace\xspace}
\newcommand{\targetNoSpace}{\ensuremath{\mathcal{T}}}
\newcommand{\target}{\targetNoSpace\xspace}
\newcommand{\adversaryPrivacyNoSpace}{\ensuremath{\mathcal{A^P}}}
\newcommand{\adversaryPrivacy}{\adversaryPrivacyNoSpace\xspace}
\newcommand{\requirement}[1]{R#1}
\newcommand{\challenge}[1]{C#1}
\newcommand{\scaleTable}[1]{\scalebox{.68}{#1}}
\newcommand{\inputdomainNoSpace}{\ensuremath{\mathcal{X}}}
\newcommand{\inputdomain}{\inputdomainNoSpace\xspace}
\newcommand{\triggersetNoSpace}{\ensuremath{\mathcal{I}}}
\newcommand{\triggerset}{\triggersetNoSpace\xspace}
\newcommand{\nClients}{\ensuremath{N}\xspace}
\newcommand{\numberOfMaliciousClients}{\ensuremath{N_{\adversary}}\xspace}
\newcommand{\cifar}{CIFAR-10\xspace}
\newcommand{\constrainandscale}{\emph{constrain-and-scale}\xspace}
\newcommand{\sota}{\mbox{state-of-the-art} }
\newcommand{\fpr}{FPR\xspace}
\begin{document}
%
\title{\paperTitle}
\author{\IEEEauthorblockN{Phillip Rieger$^*$}
\IEEEauthorblockA{Technical University of Darmstadt\\
phillip.rieger@trust.tu-darmstadt.de}
\and
\IEEEauthorblockN{Torsten Krauß$^*$}
\IEEEauthorblockA{University of Würzburg\\
torsten.krauss@uni-wuerzburg.de}
\and
\IEEEauthorblockN{Markus Miettinen}
\IEEEauthorblockA{Technical University of Darmstadt\\
markus.miettinen@tu-darmstadt.de}
\and
\IEEEauthorblockN{\hspace{3cm}Alexandra Dmitrienko}
\IEEEauthorblockA{\hspace{3cm}University of Würzburg\\
\hspace{3cm}alexandra.dmitrienko@uni-wuerzburg.de}
\and
\IEEEauthorblockN{Ahmad-Reza Sadeghi}
\IEEEauthorblockA{Technical University of Darmstadt\\
ahmad.sadeghi@trust.tu-darmstadt.de}}



%


\IEEEoverridecommandlockouts
\makeatletter\def\@IEEEpubidpullup{6.5\baselineskip}\makeatother
\IEEEpubid{\parbox{\columnwidth}{
    Network and Distributed System Security (NDSS) Symposium 2024\\
    26 February - 1 March 2024, San Diego, CA, USA\\
    ISBN 1-891562-93-2\\
    https://dx.doi.org/10.14722/ndss.2024.23233\\
    www.ndss-symposium.org
}
\hspace{\columnsep}\makebox[\columnwidth]{}}

\maketitle

\def\thefootnote{*}\footnotetext{These authors contributed equally to this work}

\renewcommand*{\thefootnote}{\arabic{footnote}}
\begin{abstract}
\noindent Federated Learning (FL) is a promising approach enabling multiple clients to train Deep Neural Networks (DNNs) collaboratively without sharing their local training data. However, FL is susceptible to backdoor (or targeted poisoning) attacks. These attacks are initiated by malicious clients who seek to compromise the learning process by introducing specific behaviors into the learned model that can be triggered by carefully crafted inputs. Existing FL safeguards have various limitations: They are restricted to specific data distributions or reduce the global model accuracy due to excluding benign models or adding noise, are vulnerable to adaptive defense-aware adversaries, or require the server to access local models, allowing data inference attacks.

This paper presents a novel defense mechanism, \ourname, that effectively mitigates backdoor attacks in FL and overcomes the deficiencies of existing techniques. It leverages clients' feedback on individual models, analyzes the behavior of neurons in hidden layers, and eliminates poisoned models through an iterative pruning scheme. \ourname employs a server-located stacked clustering scheme to enhance its resilience to rogue client feedback. The evaluation results demonstrate that \ourname achieves a 100\% True-Positive-Rate and True-Negative-Rate across various scenarios, including IID and non-IID data distributions. Additionally, \ourname withstands adaptive adversaries while preserving the original performance of protected models. To ensure confidentiality, \ourname uses a secure and privacy-preserving architecture leveraging Trusted Execution Environments (TEEs) on both client and server sides.
\end{abstract}


%

\section{Introduction}
\label{sec:intro}

\noindent Federated Learning (FL) allows multiple clients to collaboratively train a Deep Neural Network (DNN) on their private data. 
In contrast to centralized learning approaches, in FL each client trains its own DNN locally and shares only the trained parameters of the model with an aggregation server~\cite{mcmahan2017}. Thus, FL reduces concerns regarding the privacy of the clients' local data, as they never leave the respective client, which is especially important in times of increased privacy awareness, legal restrictions, and regulations~\cite{HIPAA1996,CCPA2018,GDPR2018}. FL also improves on the resource usage, as the computationally expensive training is parallelized and outsourced to the participating clients. As a result, FL has become a popular technology and is applied in various applications, including image recognition~\cite{fl_for_breast_density,sheller,sheller2018medical,Sheller2020flinmedical}, e.g., between multiple hospitals~\cite{flWithMedicalImages,fl_for_breast_density,sheller2018medical}, natural language processing (NLP), e.g., text prediction on smartphones~\cite{hard2019keypred, mcmahan2017googleGboard}, personalization~\cite{chen2018flpersonalization}, risk classification~\cite{fereidooni2022fedcri}, or threat detection in IoT networks~\cite{nguyen2019diot}.

However, outsourcing the training process to individual clients makes FL vulnerable to poisoning attacks. Here, an adversary compromises a subset of the clients and lets them submit manipulated model updates.  Such attacks can be untargeted~\cite{fang2020Local, signflippingUntargetedPoisoningFL,antidoteUntargetedPoisoningOnAnomalyDetection} or targeted (\mbox{so-called} backdoor attacks)~\cite{bagdasaryan,nguyen2020diss,shen16Auror,wang2020attack,dbabackdoor}.  In the following, we will focus on targeted poisoning attacks that are more challenging to detect (cf.~\sect\ref{sec:problem-advmodel}). These attacks cause the aggregated model to misbehave at prediction time (also called inference) if the input sample for the DNN contains a specific adversary-controlled trigger. Moreover, in FL, the clients must trust the server because several attacks have successfully inferred information about the training data from the trained parameters of a model~\cite{ganju2018property,hayes2019logan,liu2022threats,nasr2019comprehensive,pyrgelis2017knock,salem2019updates,shokri2017membership,wang2019arxivEavesdrop,labelInference}. 

The current defenses can be broadly classified into two categories: Influence Reduction (IR) approaches~\cite{andreina2020baffle,bagdasaryan,cao2021provably,mcmahan2018iclrClipping,naseri2022local,yin2018Median} and Detection and Filtering (DF) approaches~\cite{blanchard17Krum,fung2020FoolsGold,munoz19AFA,nguyen22Flame,rieger2022deepsight,shen16Auror,zhao2020shielding,kumari2022baybfed}. IR techniques aim to limit the impact of poisoned updates on the model, while DF techniques try to detect and remove the poisoned updates. These defenses employ techniques such as clipping, noising, subgroup training, distance metrics, and client-side analysis of the final predictions using the clients' local data. However, the existing defenses still face several challenges. First, they often assume that the training data of different clients are independently and identically distributed (\iid). Hence, these defenses may not work effectively when the training data of different clients differ, thus are \nonIid. Second, adaptive attackers aware of the defense mechanisms can bypass these defenses. Finally, existing defenses do not address the problem of unauthorized access to local models, allowing inference attacks. We will elaborate on related work in detail in \sect\ref{sec:sota}.

\noindent\textbf{Our goals and contributions:} We present \ourname, a backdoor-resilient and privacy-enhancing FL architecture that effectively overcomes the limitations of existing solutions. The key rationale of \ourname is to leverage a secure client-feedback-loop, where clients conduct local validation and analyze changes in the behavior of individual neurons. We define a new \clientMetricLong (\clientMetric) that enables \ourname to identify poisoned models through iterative pruning based on multiple significance tests for analyzing the models' behavior. Although the adversary might be able to inject a backdoor without affecting the final predicted class on regular data, it cannot avoid changing the behavior of at least a subset of deep-layer neurons to introduce the backdoor functionality into the DNN. To mitigate manipulated feedback from malicious clients, the server employs a multi-layer clustering scheme to aggregate the feedback of different validation clients. Utilizing the client-feedback-loop, analyzing the changes in the behavior of individual neurons, and employing stacked robust aggregation of clients' feedback enable \ourname to effectively identify poisoned models without making assumptions about the attack or data scenarios. Using the clients' data provides a comprehensive overview of the clients' training objectives, allowing to identify both, benign and poisoned models, even in scenarios where all clients' data are disjoint (\nonIid).

To mitigate the privacy risk and prevent the feedback-loop from allowing malicious clients to perform inference attacks on the received local models, \ourname leverages secure enclaves and remote attestation to prevent unauthorized access to the local models. By extending this concept to the server, we effectively solve the problem of combining privacy-preserving aggregation with backdoor mitigation. Thus, \ourname guarantees that clients' data remains confidential and cannot be inferred from the local models.

\noindent Our contributions include:
\begin{itemize}
    \item We propose \ourname, an architecture that enables secure and privacy-preserving utilization of clients' local data for local model inspection. Thus, we provide the foundation for a new class of poisoning detection algorithms. Additionally, we remove the need for trust in the aggregation server by utilizing secure, attestable enclaves on the server side and combining the backdoor detection algorithm with an efficient, secure aggregation~(\sect\ref{sec:approach-feedbackArchitecture}).
    
    \item  We design a novel backdoor detection algorithm that analyzes the \textit{hidden layer outputs} of local models to distinguish between benign and backdoored models. As the adversary's primary goal is to change the model's predictions for inputs containing the trigger, it cannot disguise the backdoor in all hidden layers without reducing the attack impact. The in-depth analysis and iterative pruning enable \ourname to effectively identify benign and poisoned updates even in \nonIid data scenarios against sophisticated, defense-adapted attacks~(\sect\ref{sec:approach-validationAlgorithm}).

    \item We conducted an extensive evaluation of the efficiency and effectiveness of \ourname on various FL scenarios to analyze different attack parameters, including various \nonIid scenarios, poisoning rates, backdoor types, and datasets such as CIFAR-10~\cite{krizhevsky2009learning} and MNIST~\cite{deng2012mnist}. By analyzing changes in the behavior of the neurons, \ourname achieved 100\% True-Positive-Rates (TPRs) and True-Negative-Rates (TNRs) in a wide range of scenarios, outperforming existing defenses while avoiding their limitations. The runtime evaluation showed an acceptable overhead of 29.5 seconds on average for the client-side validation of 20 models using enclaves on the CPU~(\sect\ref{sec:eval}).
    
\end{itemize}
We are currently integrating \ournameGen source code into the OpenFL framework~\cite{openfl_citation} to ensure that \ourname can be used not just for further research in this area but also for real-world applications~\footnote{\url{https://github.com/TRUST-TUDa/crowdguard}}.

\section{Background}
\label{sec:background}
\noindent In the following, we describe the necessary background about Federated Learning (FL) in \hyperref[sec:background:fl]{\sect\ref{sec:background:fl}} and targeted poisoning attacks on FL in \hyperref[sec:background:bdfl]{\sect\ref{sec:background:bdfl}}.

\subsection{Federated Learning}\label{sec:background:fl}
\noindent Federated Learning (FL)~\cite{federatedoptimization,mcmahan2017,surveyFLconceptandapplication} is used for generating or improving a shared machine learning (ML) model, i.e., a Deep Neuronal Network (DNN), by collaborative efforts of multiple clients $C_k \in \{C_1, \ldots C_N\}$ and a server \server in an iterative process~\cite{mcmahan2017}. The major benefit of FL is that the clients $C_k$ use their local data $\mathcal{D}_k$ for the training process. These \mbox{client-located} data do not have to be shared with the server \server. Hence, such training is more privacy-preserving than vanilla ML. Additionally, the extensive computations during learning are distributed to multiple clients, so no \mbox{cost-intensive infrastructure is needed at \server.}

The learning process takes place over multiple \textit{FL rounds} that are supervised by the server \server. At the start of each round $t$, \server first deploys a global model $G^t$ to a randomly selected group of clients $C_i \in \{C_1, \ldots,  C_n\}$, which is a subset of the total $N$ clients. The $n$ chosen clients initialize their local model $L^t_i$ with $G^t$ and continue using their local dataset $\mathcal{D}_i$ to train the new local model $L^t_i$. The training is a regular learning process configured by a number of hyper-parameters like, e.g., the learning rate, that optimizes one loss function. Afterward, \server collects all $L^t_1, \ldots, L^t_n$ models and aggregates them to a new global model $G^{t+1}$, by averaging the differences and adding them to $G^t$~\cite{mcmahan2017}.

In detail, \server computes the update of each weight for each model, builds the average of these contributions and adds the resulting value to the global model~\cite{mcmahan2017}. This algorithm is called \textit{FederatedAveraging (\fedavg)}. The final contributions are weighted with the global learning rate $\delta$ (see \hyperref[eq:1]{\equ\ref{eq:1}})~\cite{konevcny2016federated}. 
After computing a new global model $G^{t+1}$, \server can initialize a new round, e.g.,  until a round limit is reached.

\begin{equation} \label{eq:1}
    G^{t+1} = G^t + \delta (\frac{1}{n} \sum_{i=1}^{n} (L^t_i - G^t))
\end{equation}

\noindent\textbf{Client Data Distributions.}
One major point that influences the performance of any FL system regarding model accuracy but also complicates the security mechanisms against malicious contributions is the underlying distribution of the training datasets $\mathcal{D}_k \in \{\mathcal{D}_1, \ldots, \mathcal{D}_\nClients\}$~\cite{fedavgNonIIDconvergence,noniidonimages}. Even if the overall amount of samples for each label in the whole system over all clients $C_1, \ldots,  C_\nClients$ is equal, the data is unlikely to be uniformly distributed among all clients so that all $\mathcal{D}_k$ follow the same distribution~\cite{mcmahan2017}. Contrary to an \textit{independent and identically distributed (IID)} data scenario, a \textit{\nonIid} case naturally delivers divergent trained models from each client. \nonIidBig can manifest with different severities~\cite{zhu2021noniidsurvey}, i.e., all clients can have samples of all available labels, but the overall count for each label differs~\cite{fedbn}. It is common to introduce a peak \nonIid rate of $q \in [0,1]$ in sample counts for one label, the so-called \textit{main label} of the client~\cite{cao2021fltrust,nguyen22Flame}. The rest of the labels are assumed to follow a uniform distribution~\cite{cao2021fltrust,nguyen22Flame}. Alternatively, one can consider a specific distribution for all sample counts, like the Dirichlet~\cite{dirichletdistribution} or normal distribution, with the peak at the main label.

Another situation is when some clients miss one label entirely or only have data from one or two labels, which are referred to as full \textit{1-class} and \textit{2-class \nonIid}. The latter two are special cases of a uniform \nonIid distribution ($q=1$).  Most of the backdoor defenses on FL, as well as approaches improving the aggregation function, focus on either 1-class and/or 2-class \nonIid setups or the Dirichlet distribution with different main labels within the clients' datasets.

\subsection{Poisoning Attacks on FL}
\label{sec:background:bdfl}
\noindent In the past, FL has been shown to be vulnerable to so-called poisoning attacks. 
Such attacks can be untargeted~\cite{antidoteUntargetedPoisoningOnAnomalyDetection,blanchard17Krum} to decrease the accuracy of the model and reduce the convergence speed of the global model~\cite{fang2020Local,untargetedPoisoningFLFederatedVariance, signflippingUntargetedPoisoningFL}, or \textit{targeted}, also called \textit{backdoor attacks}~\cite{bagdasaryan, backdoorDataPoisoningNoise, backdoorOnNonFLModel, backdoorOnNonFLModelViaReflection, backdoorAttacksReview}, which try to add additional functionality to a model while maintaining the main task accuracy (MA). All backdoor attacks have in common that there exists an \textit{input trigger}, which is embedded within the raw data activating the backdoor. For a model that predicts labels for samples from a domain $\inputdomain$, the purpose of the backdoor is to make the poisoned model $G_*$ predicting the \textit{target label} $\target_\adversary$ when feeding a sample from the trigger set $\triggerset\subset\inputdomain$  to the model. Therefore, the adversary \adversary wants to achieve a high backdoor accuracy (BA) for samples from \triggerset, as formalized in \hyperref[eq:2]{\equ\ref{eq:2}}.

\begin{figure}[tb]
    \centering
    \subfloat[\label{fig:triggers:cartrigger}\centering Pixel]{{\includegraphics[height=2.3cm]{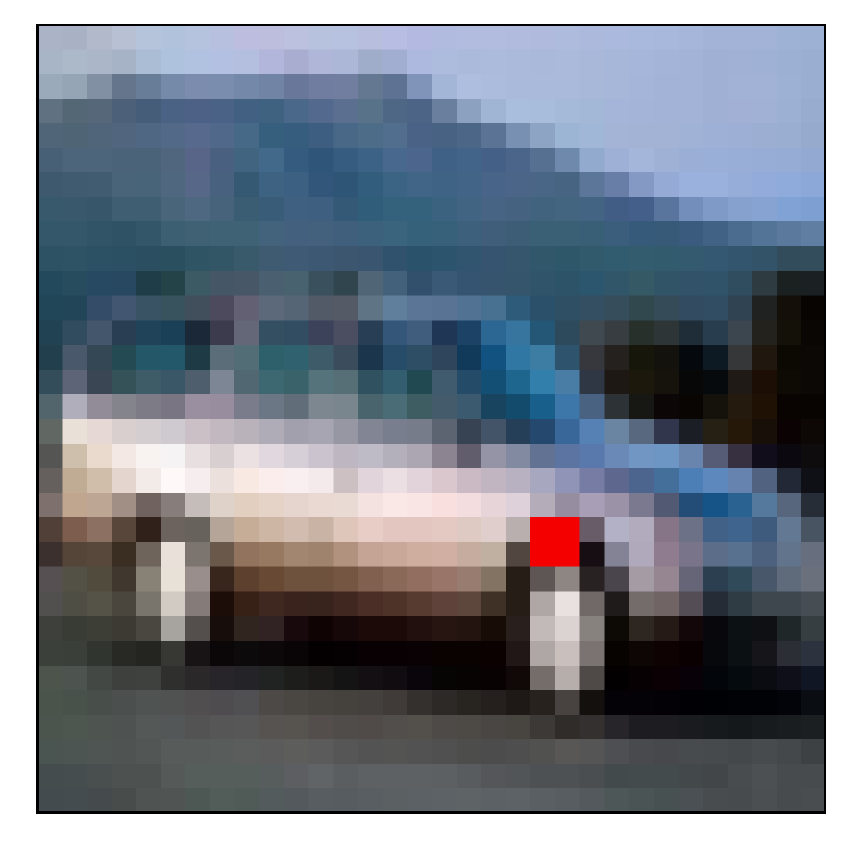} }}%
    \qquad
    \subfloat[\label{fig:triggers:carmalicious}\centering Semantic]{{\includegraphics[height=2.3cm]{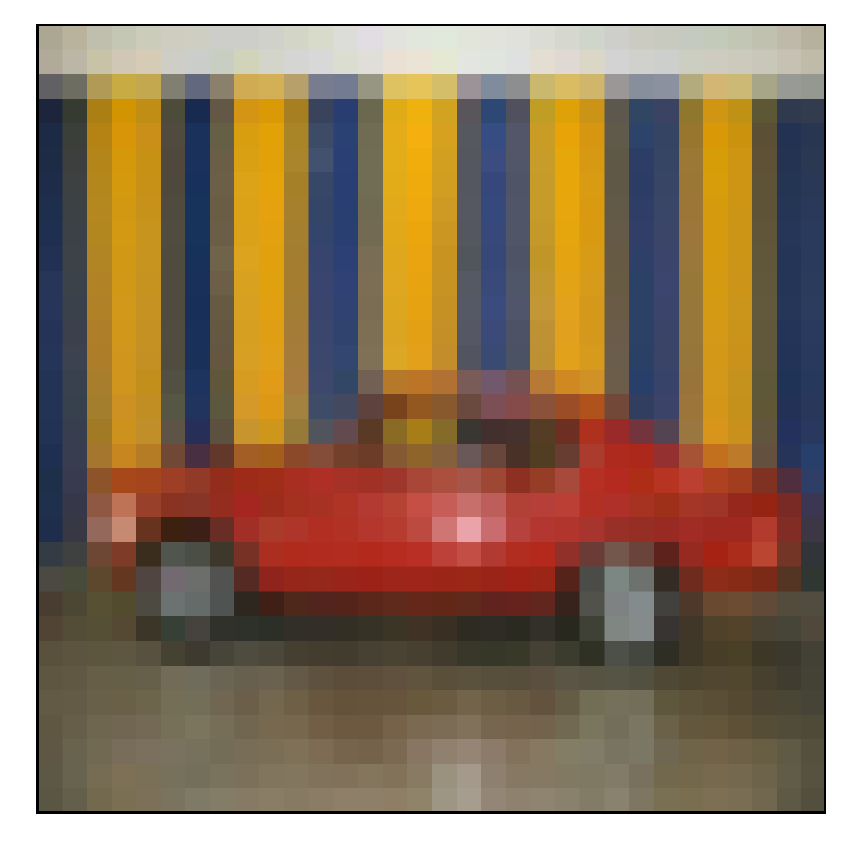} }}%
    \qquad
    \subfloat[\label{fig:triggers:car}\centering Benign]{{\includegraphics[height=2.3cm]{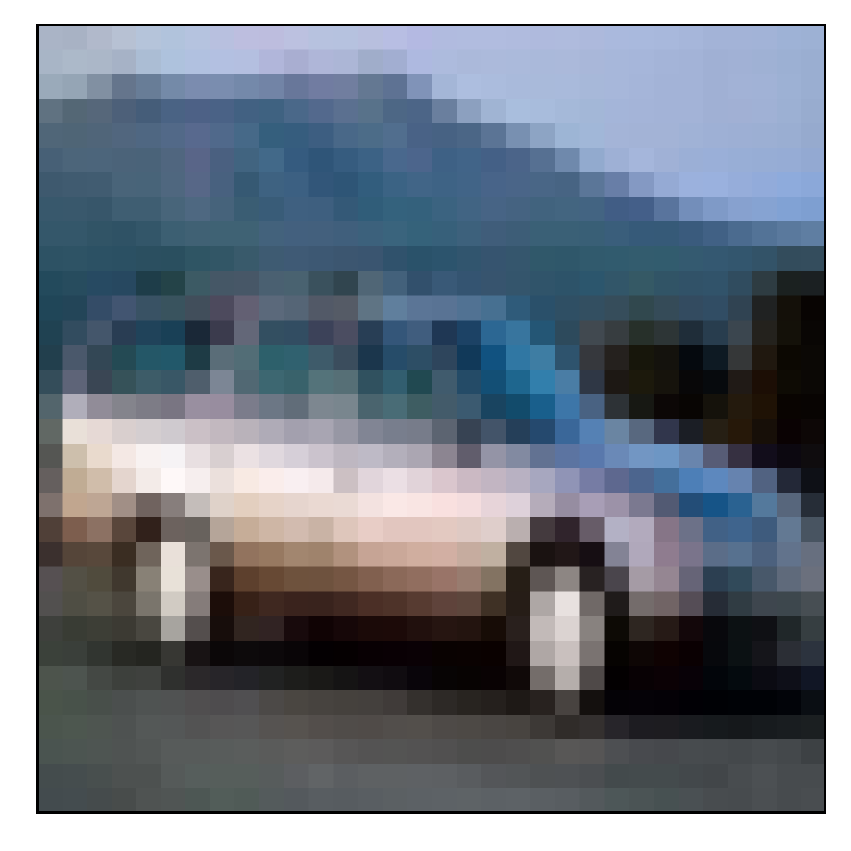} }}
    \caption{Comparison of Backdoor Triggers.}
    \label{fig:triggers}
\end{figure}

\begin{equation} \label{eq:2}
    BA = \frac{|\{x \in \triggersetNoSpace \; : \; f(x, G_*) = \target_\adversary\}|}{|\triggersetNoSpace|}
\end{equation}

Without further inspection, such backdoors remain undetected within the resulting global model and pose a danger to the model user. In FL, a poisoning attack can occur if one or more clients $C_i\in \{C_1, \ldots C_n\}$
are malicious and submit a manipulated local model to the server for compromising the aggregated model. Dependent on the attack algorithm and the attacker's capabilities, the adversary can manipulate the input data, the complete learning process including hyper-parameters, and the final weights of the trained local models to inject and hide a backdoor (with high BA while maintaining high MA). Furthermore, it can adapt to the defense by imitating the behavior of benign clients (e.g., scaling of the model weights) to stay undetected but still effective. 

Different triggers have been proposed to activate the backdoors: 1) \textit{Pixel backdoors} that get activated by a certain pixel pattern, like a red rectangle~\cite{bagdasaryan,badnets,trojanTriggerTargeted}, which can also be distributed in fractions over multiple poisoned local models~\cite{dbabackdoor}, 2) a \textit{Label-Swap} backdoor that mislabels all samples of one class to a target class, or 3) a \textit{Semantic Backdoor} that is activated when certain characteristics, e.g., a car in front of a striped background, is present in the input~\cite{bagdasaryan}. Examples of all triggers are shown in \hyperref[fig:triggers]{\fig\ref{fig:triggers}}. We elaborate on these triggers in \hyperref[app:backdoors]{\app\ref{app:backdoors}}. To achieve his goal, the adversary \adversary chooses one or more of the following concepts:\\
\noindent\textbf{Data Poisoning:} \adversary manipulates the training dataset $\mathcal{D}_i$, so that the resulting $\mathcal{D}^{\adversary}_i$ includes samples containing the trigger. To trade-off the effectiveness against the detectability of the attack, a respective ratio of malicious to benign samples, the so-called \textit{Poison Data Rate (PDR)}, must be chosen by~\adversary.\\
\noindent\textbf{Model Poisoning:} The adversary manipulates the training algorithm itself by changing hyper-parameters. Additionally, \adversary can optimize against several objectives~\cite{multipleGradDecend} in the form of additional loss functions (weighted by an $\alpha$ parameter) and constrain the loss(es) to stay as close as possible to benign behavior, especially if the adversary is aware of the defense. The weights of the resulting trained local models can be adapted to benign models to circumvent straightforward defenses that analyze the weights of all contributions and remove extreme outliers. This process conducted by an adaptive attacker is called \textit{constrain-and-scale}~\cite{bagdasaryan}.

\subsection{Trusted Execution Environments}
\noindent Trusted Execution Environments (TEEs) are programmable secure areas located within a processor that allow the execution of applications within secure enclaves, isolated from the remaining system. By using, e.g., memory encryption, access to them from outside the enclave, even from privileged processes, is restricted and thus guarantees the confidentiality of the enclaves' data. Attestation allows to verify the authenticity of the TEE and the integrity of code executed within the TEE~\cite{costan2016intel}. Examples of TEEs are Intel SGX~\cite{costan2016intel}, AMD SEV~\cite{kaplan2016amd}, ARM TrustZone~\cite{armtrustzone}, \mbox{and Nvidia Confidential Computing~\cite{nvidiaH100}.}

\section{Problem Setting}
\label{sec:problem}

\noindent In the following, we describe the considered system (\hyperref[sect:problem:system]{\sect\ref{sect:problem:system}}) and characterize the threat model (\hyperref[sec:problem-advmodel]{\sect\ref{sec:problem-advmodel}}).

\subsection{System Setting}
\label{sect:problem:system}
\noindent In the setup, we consider mainly cross-silo settings\footnote{Real-world scenarios and projects are, e.g. the FeTS project~\cite{fets}.} with \nClients clients $C_1, \ldots C_\nClients$ that have private datasets $D_1, \ldots, D_\nClients$ but will not share any data to prevent privacy leakages~\cite{fets}. Further, we consider an aggregation server \server that receives the individual models and aggregates them using \fedavg~\cite{mcmahan2017}. Aligned with recent work on poisoning attacks~\cite{andreina2020baffle,bagdasaryan,nguyen22Flame}, we use an adapted version of the regular \fedavg algorithm and scale the local models' contributions equally with $\nicefrac{1}{n}$ instead of weighting them based on their dataset sizes. This prevents malicious clients from artificially increasing their impact by reporting wrong dataset sizes. We keep the global learning rate constant at $\delta = 1$. In principle, multiple aggregation algorithms exist~\cite{blanchard17Krum,bulyan,munoz19AFA,yin2018Median}, which either provide better performance or are more robust against byzantine contributions from local models. Our method can be used with different aggregation techniques since it is applied \mbox{before the aggregation takes place.}

We assume that each client and the server have an arbitrary TEE available, allowing the execution of code in the secure enclaves while isolating code and memory from the remaining system, including privileged parts. Thus, the TEEs shall prevent the remaining system from learning the data inside the enclave and therefore preserving the data's confidentiality. Further, the TEE needs to allow a remote machine to attest the code of the executed enclave. Depending on the application, e.g., in cross-silo applications where different institutions like hospitals collaboratively train a DNN, the machines performing the local training can be assumed to be powerful platforms, providing standard hardware features like TEEs and thus making this assumption reasonable.

\begin{figure}[tb]
    \centering
    \includegraphics[width=.9\columnwidth]{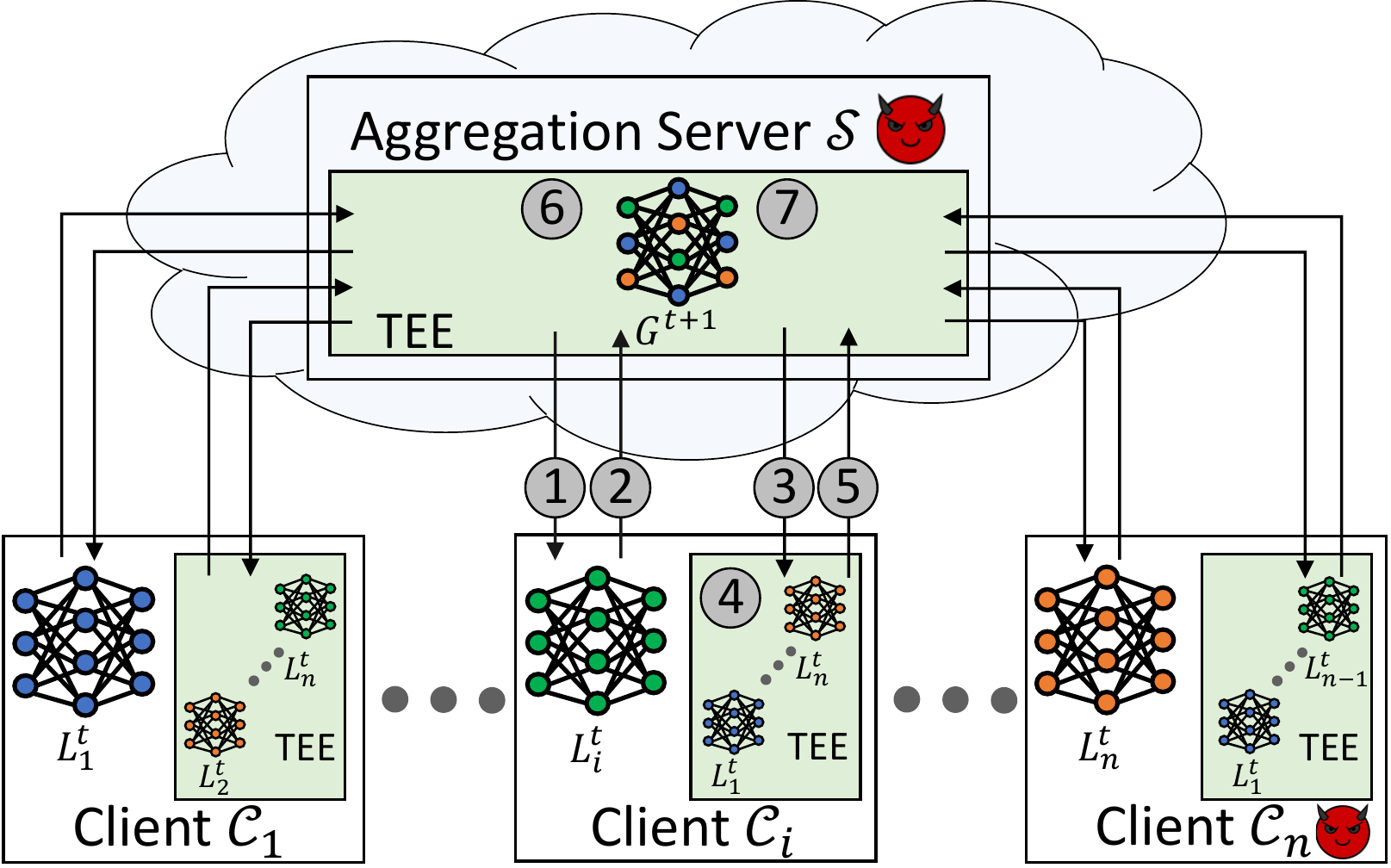}

    \caption{Overview and steps of \ourname.}
    \label{fig:overview}
\end{figure}

An overview of the considered system is shown in \hyperref[fig:overview]{\fig\ref{fig:overview}}, showing the clients, the aggregation server, and the individual TEEs (marked in green). \hyperref[fig:overview]{\fig\ref{fig:overview}} also shows the individual steps of our scheme, which we will discuss in \sect\ref{sec:problem-design}.

In contrast to existing work, we do not make any assumptions about the data distributions. Thus, the individual clients' data can follow the same distribution (\iid), be distributed differently (\nonIid), or even be disjoint. 

\subsection{Adversary Model}\label{sec:problem-advmodel}
\noindent We consider two adversaries. The first, \adversary, aims to inject a backdoor into the FL system, while \adversaryPrivacy aims to learn information about the clients' data from the local model updates, violating the privacy of the data.

\subsubsection{Poisoning Attacker \adversary}
$\mathcal{A}^B$ (for the sake of brevity, denoted as \adversary) aims to manipulate the model that is resulting from the FL process and injects a backdoor into it by utilizing data and/or model poisoning (see \hyperref[sec:background:bdfl]{\sect\ref{sec:background:bdfl}}). If a certain, adversary-chosen trigger is present in the input (cf. \sect\ref{sec:background:bdfl}), the backdoor shall make the aggregated model $G^{t+1}$ predicting an adversary-chosen target class $\target_{\adversary}$. From this goal, two objectives follow:

\noindent\textbf{O1 - Attack Impact:} To inject a backdoor successfully, \adversary aims to make the aggregated model $G^{t+1}$ predicting the backdoor target class $\target_{\adversary}$ for all trigger samples \triggerset from the input domain \inputdomain. Thus, its objective is to maximize the accuracy on the backdoor task, e.g., the BA. If the server \server notices the attack, it repeats the training process with a subset of clients until no backdoor is noticed anymore or filters the poisoned contributions. Therefore, a second objective for the adversary \adversary is:

\noindent\textbf{O2 - Stealthiness:} Make the poisoned model updates inconspicuous such that \server can neither identify the poisoned updates nor notices the performed backdoor attack\footnote{This differs backdoor from untargeted attacks, as the latter one can always be identified by a drop in the models' utility}.

From O2 also follows that the attack \textit{must not} reduce the performance of the aggregated model on the main task (MA). If the predictions of the aggregated model $G^{t+1}$ for a sample $x\in\mathcal{X}$ are denoted as $f(x, G^{t+1})$, $G^{t+1}$ is the aggregated model without the poisoning attack, and $G_*^{t+1}$ is the aggregated model including the poisoned contributions, then O1 and O2 result in the following goal of the adversary \adversary: 
\begin{equation}
    f(x, G_*^{t+1}) = \begin{cases}
\target_{\adversary} \text{ if } x\in \mathcal{I}\\
f(x, G^{t+1}) \text{ if } x\not\in \mathcal{I}
\end{cases}
\end{equation}

Aligned with previous work~\cite{blanchard17Krum,munoz19AFA,nguyen22Flame,shen16Auror,li2023flairs}, we assume that \adversary fully controls $n_{\adversary} < \nicefrac{n}{2}$ clients in one round and overall $\numberOfMaliciousClients < \nicefrac{\nClients}{2}$ clients in the whole FL system.\footnote{In each round $n$ clients are selected for training out of all $N$ clients.} Thus, it can freely manipulate their local datasets $D_i$, change the training process, or even manually change the submitted parameter updates and replace parameters with arbitrary numbers\footnote{It should be noted that an adversary can manipulate the local dataset that is given to the client-side enclaves to manipulate their behavior.}. The ratio of poisoned models to all local models $\nicefrac{n_{\adversary}}{n}$ is denoted as Poisoned Model Rate (PMR). Further, we assume that \adversary knows all algorithms the server or clients execute. Thus, \adversary can adapt its attack strategy and the client's behavior, e.g., hyper-parameters and the objective of the local training, with respect to the deployed defense to make the attack inconspicuous (adaptive adversary).

\subsubsection{Privacy Attacker \adversaryPrivacy}
The second adversary, \adversaryPrivacy, aims to reconstruct information about the clients' local data. Aligned with existing work~\cite{andreina2020baffle,khazbak2020mlguard,nguyen22Flame}, we consider only privacy attacks that learn information about the clients' data by analyzing the local model updates. The aggregation of FL anonymizes the individual contributions, preventing \adversaryPrivacy from associating gained information with a specific client, and also smoothens the parameters. Thus, we will consider privacy attacks on the aggregated model out of the scope of this work.

In our threat model, we consider \adversaryPrivacy to be a malicious attacker that has arbitrary control over the aggregation server. Further, \adversaryPrivacy can control some of the clients to analyze any other client's local model that this client might receive. In contrast to existing work, the considered adversary \adversaryPrivacy even fully controls the server \server. However, the benign clients and the server can use remote attestation to verify the code and authenticity of the secure enclaves that are running on the server and the clients, respectively, before sharing local models. 

\subsubsection{TEE Security Assumptions}
In the following, we consider arbitrary TEEs that isolate executed secure enclaves and allow a remote machine to attest the running enclave (cf.~\sect\ref{sect:problem:system}). Thus, \ourname is not restricted to TEEs of certain manufacturers. However, we assume that all used TEEs are trusted. Therefore, attacks on the used cryptographic algorithms and attacks that extract keys burned into the TEE are out of the scope of this paper.

Recently, several side-channel attacks have been proposed that extract data, e.g., the received models or cryptographic keys, from TEEs~\cite{sidechannelsgxusenix,huang2021aion,wang2018interface}.  As discussed in \hyperref[app:sidechannels]{\app\ref{app:sidechannels}}, with this model stealing or inference attacks could be executed. There exist already works to counter such attacks~\cite{barthe2020formal,daniel2020binsec,brasser2019dr,sang2022pridwen}. Therefore, we consider attacks on the TEE architecture to be out of the scope of this work.

\subsection{Requirements and Challenges}
\label{sec:prob:challanges}
\noindent Based on the characterization of \adversary and \adversaryPrivacy, the following requirements for a backdoor defense can be derived:\\
\noindent\textbf{\requirement{1}:} Prevent the backdoor attack, i.e., $\forall x\in\mathcal{I}: \; f(x, G^{t+1}_*)=f(x, G^{t+1})$. \\
\noindent\textbf{\requirement{2}:} To be practical, the defense scheme must not reduce the benign performance of the resulting FL model, especially in the absence of any attack. Therefore, if no attack was performed 
and $G^{t+1}$ is the aggregated model obtained using \ourname, while $\hat{G}^{t+1}$ was obtained using plain \fedavg, 

then both model's outputs should be equal: $\forall x\in \inputdomain: \; f(x, \hat{G}^{t+1})=f(x, G^{t+1})$. \\
\noindent\textbf{\requirement{3}:} The defense must preserve the clients' privacy. Thus, the server \server must not be able to access the models for running inference attacks. Nor should any other party, e.g., the clients, be able to run inference attacks on the models of other clients.

From these requirements, a number of challenges follow that \ourname will address in the rest of the paper:

\noindent\textbf{\challenge{1}:} How to effectively distinguish benign and poisoned models, especially for \nonIid scenarios, to fulfill \requirement{1}? \hyperref[sec:approach]{\sect\ref{sec:approach}} explains how \ourname can distinguish poisoned models and benign models, being trained on abnormal data.\\
\noindent\textbf{\challenge{2}:} The server \server must not be able to access the individual local models as this would enable \server to run inference attacks (cf. \requirement{3}). However, to identify poisoned model updates, \server has to inspect the model updates $L^{t}_i$. A challenge that \ourname will address is, therefore, how to inspect the local models without enabling any party to extract knowledge from them.\\
\noindent\textbf{\challenge{3}:} \ourname uses the predictions, including the hidden state outputs, of the local models on the local data of other clients for identifying a backdoor. However, the backdoor attack should not change the predictions for non-triggered input samples (cf. O2). Since it is unlikely that benign clients have many triggered input samples, a challenge that \ourname will solve is how to use clients' local data for identifying the backdoor without having triggered samples.

\subsection{Design of \ourname}
\label{sec:problem-design}
\noindent Due to the absence of validation data, existing approaches for backdoor mitigation on the aggregation server are restricted to using vector metrics or outlier detection~\cite{munoz19AFA,shen16Auror,nguyen22Flame}. These methods have limited effectiveness in \nonIid settings or against sophisticated adversaries (cf. \sect\ref{sec:sota}). To overcome these limitations, \ourname involves sending the local models to the clients and collecting their feedback on the individual models to identify backdoored models. The use of such a client feedback-loop and a validation algorithm that analyzes changes in the outputs of the models' individual layers enables \ourname to effectively identify poisoned models even in \nonIid settings. \hyperref[fig:overview]{\fig\ref{fig:overview}} provides an overview of the individual steps of our approach, which are depicted in more detail in \hyperref[app:overview]{\app\ref{app:overview}}. To protect the local models' confidentiality, all transmissions are encrypted, the code runs in TEEs, and each enclave is attested before receiving any model.

Each client begins with the setup of its secure-enclave, before receiving the global model from the server, trains its local model\footnote{Depending on the respective application scenario, there are reasons for performing also the training inside a TEE, such as confidentiality of the training data, while there are also reasons against it, e.g., the computational overhead. \ourname focuses on a backdoor resilient aggregation and supports both operational modes.}, and sends the model encrypted to an enclave running on the aggregation server \server (steps 1 and 2 in \hyperref[fig:overview]{\fig\ref{fig:overview}}). After collecting the individual model updates, \server sends the local models to secure enclaves running on the clients $C_i\in {C_1, \ldots, C_n}$, which are utilized as validation clients (step 3 in \hyperref[fig:overview]{\fig\ref{fig:overview}}). The validation is performed in a secure enclave to ensure confidentiality and prevent the client feedback-loop from increasing the surface for privacy attacks. Then, the client-side enclaves validate the models by analyzing changes in hidden layer outputs using the local datasets $\mathcal{D}_i \in {\mathcal{D}_1, \ldots, \mathcal{D}_n}$ (addressing \challenge{1}), denoted as Step 4 in \hyperref[fig:overview]{\fig\ref{fig:overview}}. Our method employs a novel metric called \clientMetricLong (\clientMetric) to analyze the obtained values iteratively and identify poisoned models based on statistical significance tests. The clients provide feedback to the server by voting for each model, indicating whether the local model is benign or suspicious (step 5 in \hyperref[fig:overview]{\fig\ref{fig:overview}}). Afterward, the server applies a stacked clustering schema to combine the votes provided by the clients (step 6 in \hyperref[fig:overview]{\fig\ref{fig:overview}}). This step mitigates manipulated votes from malicious clients who may provide manipulated data to the client-side enclave. Finally, the server removes the models marked as poisoned and uses a configured aggregation rule to aggregate the remaining models before sending the aggregated model $G^{t+1}$ back to the clients for further training rounds (steps 6 and 7 in \hyperref[fig:overview]{\fig\ref{fig:overview}}). It should be noted that although we focus on \fedavg as the aggregation rule in this paper, other rules such as Krum~\cite{blanchard17Krum}, trimmed mean~\cite{yin2018Median}, or median~\cite{yin2018Median} can be used.

\section{\ourname}
\label{sec:approach}
\noindent In the following, we describe the details of \ourname, starting with the overall architecture that allows utilizing clients' feedback in \hyperref[sec:approach-feedbackArchitecture]{\sect\ref{sec:approach-feedbackArchitecture}}. Afterward, in \hyperref[sec:approach-validationAlgorithm]{\sect\ref{sec:approach-validationAlgorithm}}, we describe the algorithm that is executed on the client-side of \ourname to create the feedback and in \hyperref[sec:approach-voting]{\sect\ref{sec:approach-voting}}, how the feedback of different clients is aggregated to be robust against manipulated feedback of malicious clients.

\subsection{Privacy-Enhancing Architecture for Clients' Feedback}
\label{sec:approach-feedbackArchitecture}
\noindent Given the impracticality of assuming the aggregation server to posses validation data~\cite{rieger2022deepsight}, existing defenses are restricted to applying vector metrics, conducting outlier detection, or making predictions on randomly generated data that do not produce meaningful predictions. In addition, clients are reluctant to share their data with the server, since this would undermine the privacy advantage of FL, which allows for model training without data sharing. Secondly, transmitting large datasets to the server incurs significant communication overhead. However, sending models to the clients is also undesirable due to the risk of inference attacks by malicious clients (\adversaryPrivacy)~\cite{hayes2019logan,liu2022threats,nasr2019comprehensive,pyrgelis2017knock,salem2019updates,shokri2017membership,labelInference}. To overcome this dilemma, we propose a secure feedback loop that enables client-side validation of local models within TEEs (addressing \challenge{2}). Additionally, server-side operations are also executed within a secure enclave to prevent abuse of the clients' feedback, e.g., to obtain information about their data. This guarantees the confidentiality of the processed data, including the models and corresponding layer outputs, even if $\adversaryPrivacy$ has kernel privileges (fulfilling \requirement{3}). Attesting the enclaves prior to any data transmission guarantees that the executed code does not leak the received models (cf. \challenge{2}).

\noindent\textbf{Setup Phase.} 
During the setup phase, each client starts the secure enclave responsible for client-side validation and feedback provision to the server. Once the enclave is launched, the client shares its private dataset with the local enclave. Next, the server attests the integrity and authenticity of client-side enclaves.  To mitigate the risk of fake validation requests from the server, which could potentially exploit clients' feedback to infer their data, clients also perform an attestation process to verify the correct execution of the FL server code. Notably, the relocation of the server to a secure enclave establishes a secure aggregation scheme. Hence, clients also attest the server-enclave during the setup phase.

\noindent\textbf{Validation Phase.} 
After completing the system setup, the FL process is initiated and the clients train their local models orchestrated by the server (represented by steps 1 and 2 in \hyperref[fig:overview]{\fig\ref{fig:overview}}). After receiving the locally trained model from each client, the server distributes these models to the client-side validation enclaves (step 3 in \hyperref[fig:overview]{\fig\ref{fig:overview}}). Each enclave utilizes its local datasets to identify any models that have been poisoned (step 4 in \hyperref[fig:overview]{\fig\ref{fig:overview}}, for details on this algorithm, see \sect\ref{sec:approach-validationAlgorithm}) and submits this feedback to the server (step 5 in \hyperref[fig:overview]{\fig\ref{fig:overview}}). It is important to note, that although the code of the validation enclaves is attested, the malicious clients can still manipulate their own validation result by providing manipulated validation data to the enclave. Hence, using secure enclaves on the client side only guarantees the models' confidentiality but not the votes' integrity.

To address this issue, a robust aggregation algorithm needs to be deployed on the server side to aggregate the feedback and remove manipulated feedback as well as noisy feedback from benign clients (see \hyperref[sec:approach-voting]{\sect\ref{sec:approach-voting}}, step 6 in \hyperref[fig:overview]{\fig\ref{fig:overview}}). Finally, the server uses the aggregated feedback to remove poisoned models, aggregate the remaining models and proceed with the next FL round (step 7 in \hyperref[fig:overview]{\fig\ref{fig:overview}}).

\subsection{Hidden-Layer Analysis for Backdoor Detection}
\label{sec:approach-validationAlgorithm}
\noindent To identify poisoned models using benign validation data, illustrated as step 4 in \hyperref[fig:overview]{\fig\ref{fig:overview}}, \ourname analyses the outputs of the individual layers of the DNN to distinguish between benign and backdoored (local) models. The result is provided as votes to the server afterward. This happens in two steps: 1) extraction of a novel metric, the \clientMetricLong (\clientMetric), from the local models to inspect them for backdoors and 2)~analyzing the \clientMetric via probabilistic tests to produce voting decisions.

\noindent\textbf{\clientMetric Motivation}
Analyzing two metrics with statistical tests enables \ourname to detect different model manipulation strategies. Vectors (i.e., DNN’s parameters) can be manipulated in two ways: Without changing direction, which is the orientation of the vector (checked by Euclidean distance) or with changing direction (increased Cosine distance) to the previous vector state. Using both metrics enables the detection of malicious model changes (addressing \challenge{3}). \clientMetric carves out significant changes in the plain values of both distances. The calculation of the \clientMetric is explained in the following.

\noindent\textbf{\clientMetric Matrix Generation}
As depicted in \hyperref[alg:hlbim]{\alg\ref{alg:hlbim}} \mbox{lines 1-6}, the global model $G^t$, the local models $L^{t}_i \in \{L^{t}_1, \ldots, L^{t}_n\}$, and the validation client's local data $D_j$ are used to generate two \clientMetric matrices based on Cosine (\clientMetricNoSpace$^C$) and Euclidean (\clientMetricNoSpace$^E$) distances. Both matrices are based on the deep layer outputs (DLOs), which can be obtained by feeding the local data into both, the local models and the global model. During inference, we keep track of the outputs of each sample, each model, and each layer, depicting one DLO in a respective matrix (\mbox{lines 7-15}). Notably, \ourname analyzes all layers' DLOs. Otherwise, if \ourname only considered a subset of layers, the adversary could utilize the unconsidered layers for injecting the backdoor without being detected. To prevent such an attack, the DLO matrices are based on all layers.

\begin{algorithm}[tb]
	\caption{\clientMetric Matrix Generation for client $C_j$}
	\label{alg:hlbim} 
	\scalebox{0.85}{
	\begin{minipage}{1.18\columnwidth}
	\small
	{
	\begin{algorithmic}[1]

		\State \textbf{Input:}\\
		$G^t$, \Comment{Global model of round $t$}\\
		$L^{t}_i$, \Comment{All local contributions of round $t$ including $L^{t}_j$}\\
		$D_j$ \Comment{Local dataset of client $C_j$}
		
		\State \textbf{Output:}\\
		\clientMetricNoSpace$^{C/E}_{m_j\;m\;l}$ \Comment{\clientMetric matrices for Cosine \& Euclidean distances}
		
		\State \Comment{Generate deep layer outputs}
		\State DLO\_local$_{s\;m\;l}$ $\gets$ \{\}
		\State DLO\_global$_{s\;l}$ $\gets$ \{\}
		
		\For{$s$ in $D_j$}
		    \For{$m$ in $L^{t}_i$}
		        \State DLO\_local$_{s\;m\;l}$ $\gets$ deep\_layer\_outputs($s$, $m$)
		    \EndFor
		    \State DLO\_global$_{s\;l}$ $\gets$ deep\_layer\_outputs($s$, $G^t$)
		\EndFor
		
		\State \Comment{Distance Generation}
		\For{dist$^{C/E}$ in [COSINE-distance; EUCLIDEAN-distance]}
		    \State DLO\_dist$^{C/E}_{s\;m\;l}$ $\gets$ dist$^{C/E}$(DLO\_local$_{s\;m\;l}$, DLO\_global$_{s\;l}$)
		    \State \Comment{Scale relative distances to \clientMetric}
		    \For{dlo$^{C/E}_{s\;m\;l}$ in DLO\_dist$^{C/E}_{s\;m\;l}$}
		    \State dlo\_rel$^{C/E}$ $\gets$ dlo$^{C/E}_{s\;m\;l}$ $/$ DLO\_dist$^{C/E}_{s\;m_j\;l}$
		    \State DLO\_squared$^{C/E}_{s\;m_j\;m\;l}$ $\gets$ $|$dlo\_rel$^{C/E}$-1$|$ $*$ (dlo\_rel$^{C/E}$-1)
		    \EndFor
		    \State DLO\_avg$^{C/E}_{lab\;m_j\;m\;l}$ $\gets$ AVG(labels $lab$, DLO\_squared$^{C/E}_{s\;m_j\;m\;l}$)
		    \State \clientMetricNoSpace$^{C/E}_{m_j\;m\;l}$ $\gets$ CONCAT(labels $lab$, DLO\_avg$^{C/E}_{lab\;m_j\;m\;l}$)
		\EndFor
	\end{algorithmic}}
\end{minipage}}
\end{algorithm}

The two DLO matrices for Euclidean and Cosine distances are then used to calculate the final \clientMetric within five steps (\mbox{lines 18-25}):  1)~Distances between the global models and each local model are computed for each DLO (line 18). In these distances, the backdoor behavior is not yet detectable with high significance. 2)~A ratio of the DLOs is generated, using the validating client's local model as a reference to highlight differences between the models regarding a reference model (line 21). The rationale is that each client assumes his own model to be benign, while the ratio highlights already small differences between the model updates. If the DLOs are equal, the ratio will be one. 3)~To further highlight the differences between the ratios, the values are scaled by subtracting one and squaring the result while retaining the sign (line 22). 4)~The resulting DLO matrices are averaged over the sample dimension for each label, thereby carving out the effect for each label class separately, with the first matrix index changing from sample $s$ to label $lab$ (line 24). 5)~To reduce the matrix dimension without losing information and thus saving computational costs in subsequent stages, the $lab$ dimension is flattened by concatenating the values (line 25). It is important to keep the values from different labels separate until the pruning. This separation ensures that only samples from the same label, which traverse the model similarly, are used by our method to identify abnormal behavior. Averaging beforehand would lead to the loss of important information and limit the effectiveness of the metric.
\noindent\textbf{Voting Decision via Model Pruning}
To detect poisoned models, the \clientMetric matrices must be analyzed.\footnote{Euclidean and Cosine \clientMetric matrices are analyzed independently.} As described in \hyperref[alg:prune]{\alg\ref{alg:prune}}, \ourname first leverages a Principal Component Analysis (PCA) on the \clientMetric matrix\footnote{The first dimension  $m_j$ of \clientMetricNoSpace$^{C/E}_{m_j\;m\;l}$ is fixed to index $j$ of the validating client $V_j$. Therefore the matrix is two dimensional.} to highlight the differences between different models (cf.~\mbox{line 14}), before iteratively pruning the poisoned models.

\begin{figure*}[tb]
    \centering
    \subfloat[\label{fig:significance:0}\centering Pruning round 1]{{\includegraphics[height=2.7cm,trim={0 0.325cm 0 .2cm},clip]{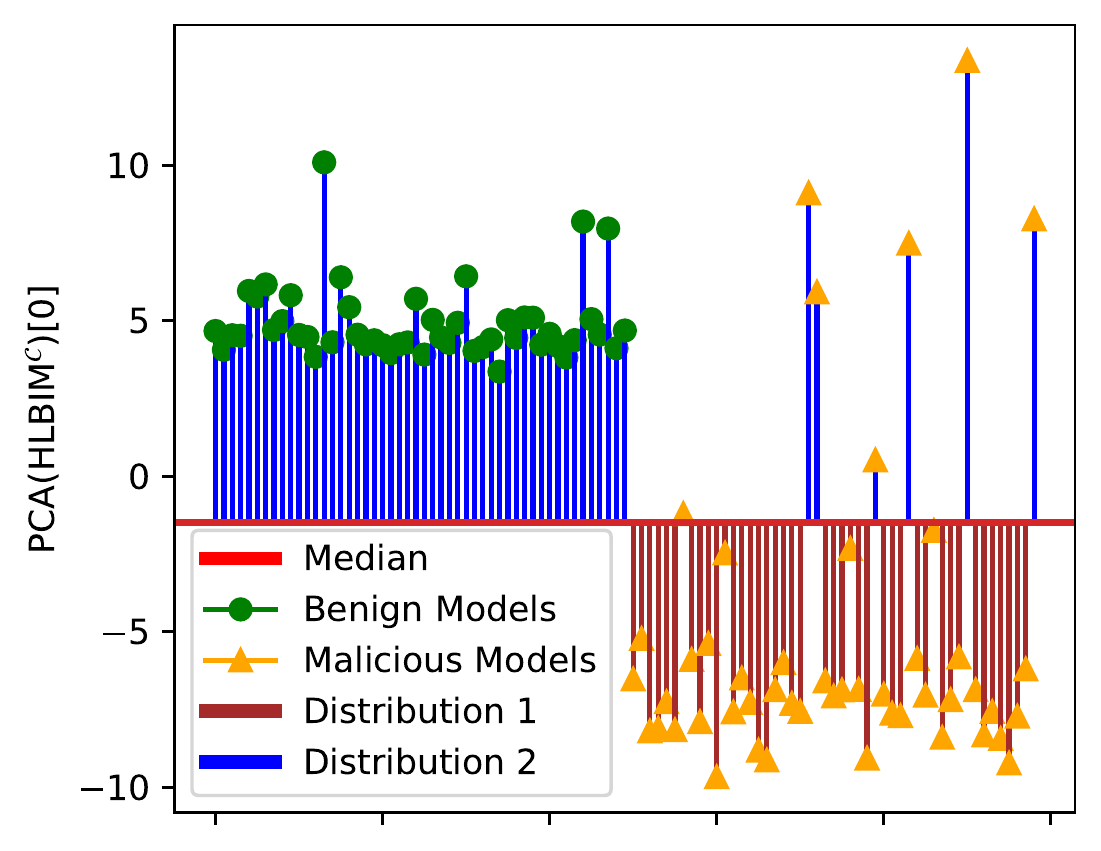} }}%
    \qquad
    \subfloat[\label{fig:significance:1}\centering Pruning round 2]{{\includegraphics[height=2.7cm,trim={0 0.325cm 0 .2cm},clip]{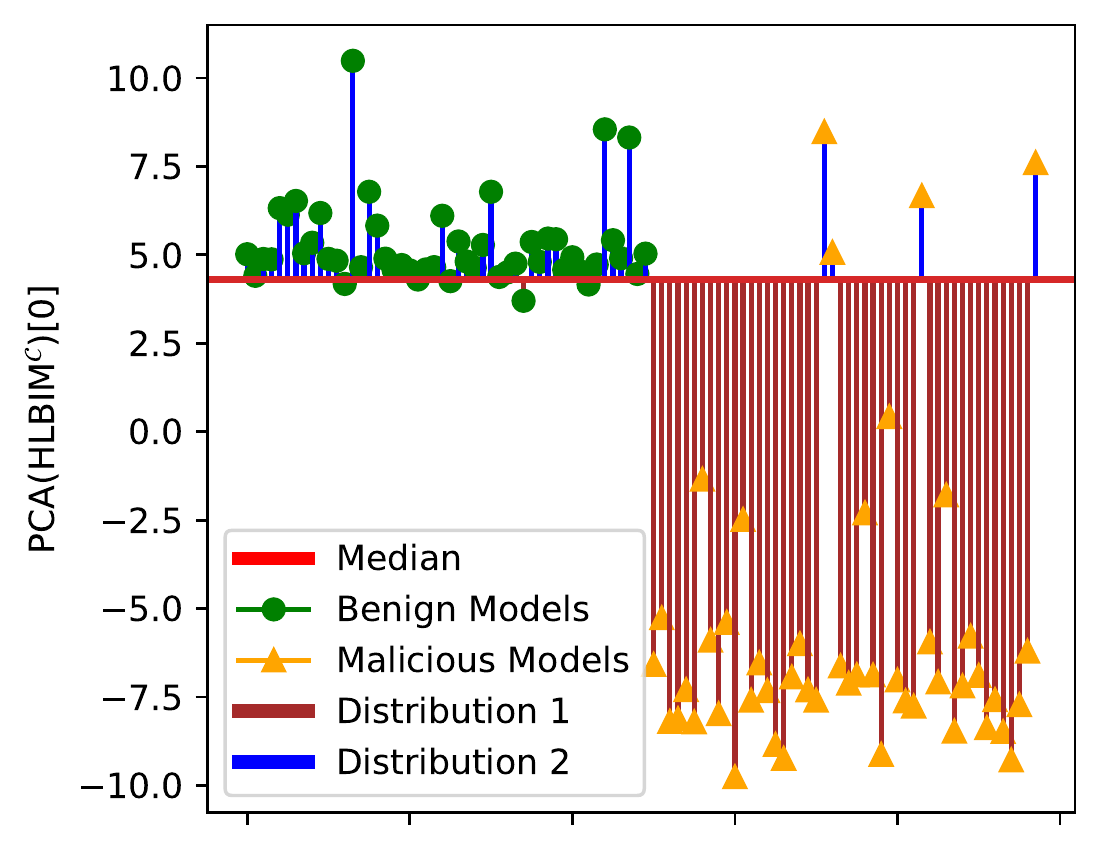} }}%
    \qquad
    \subfloat[\label{fig:significance:2}\centering Pruning round 3]{{\includegraphics[height=2.7cm,trim={0 0.325cm 0 .2cm},clip]{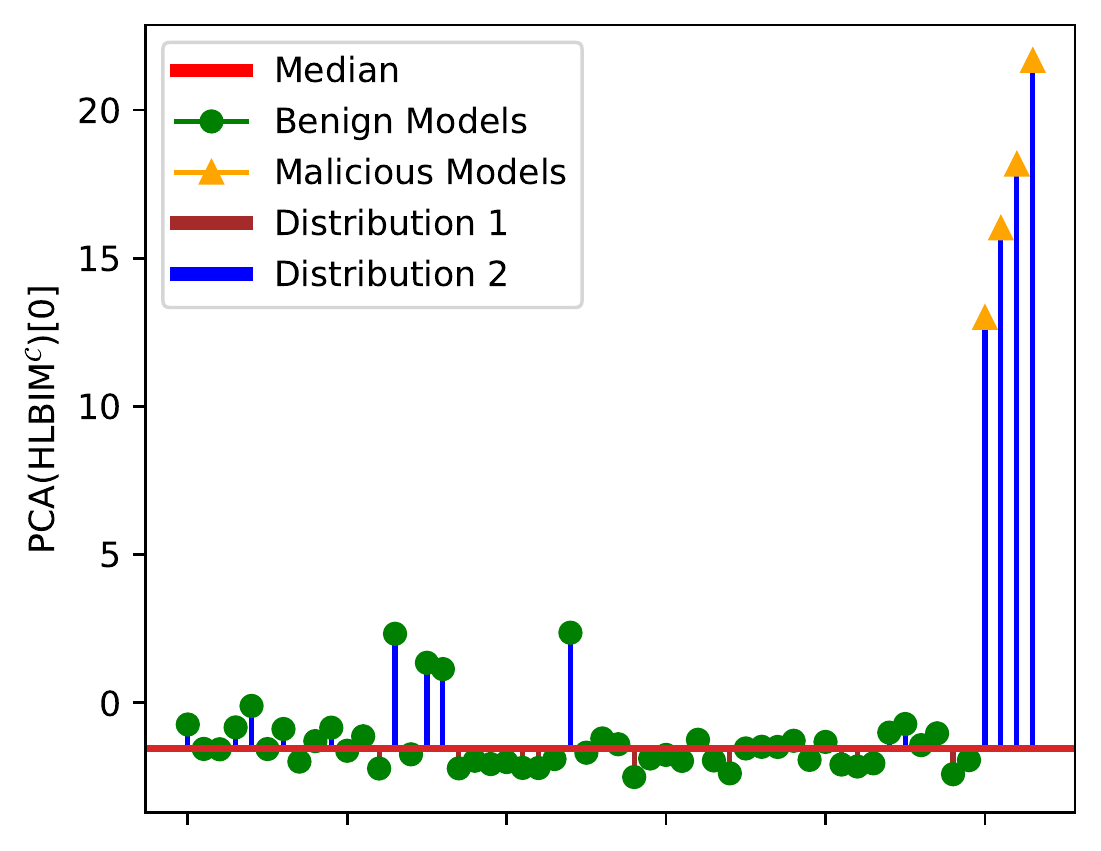} }}%
    \qquad
    \subfloat[\label{fig:significance:3}\centering Pruning round 4]{{\includegraphics[height=2.7cm,trim={0 0.325cm 0 .2cm},clip]{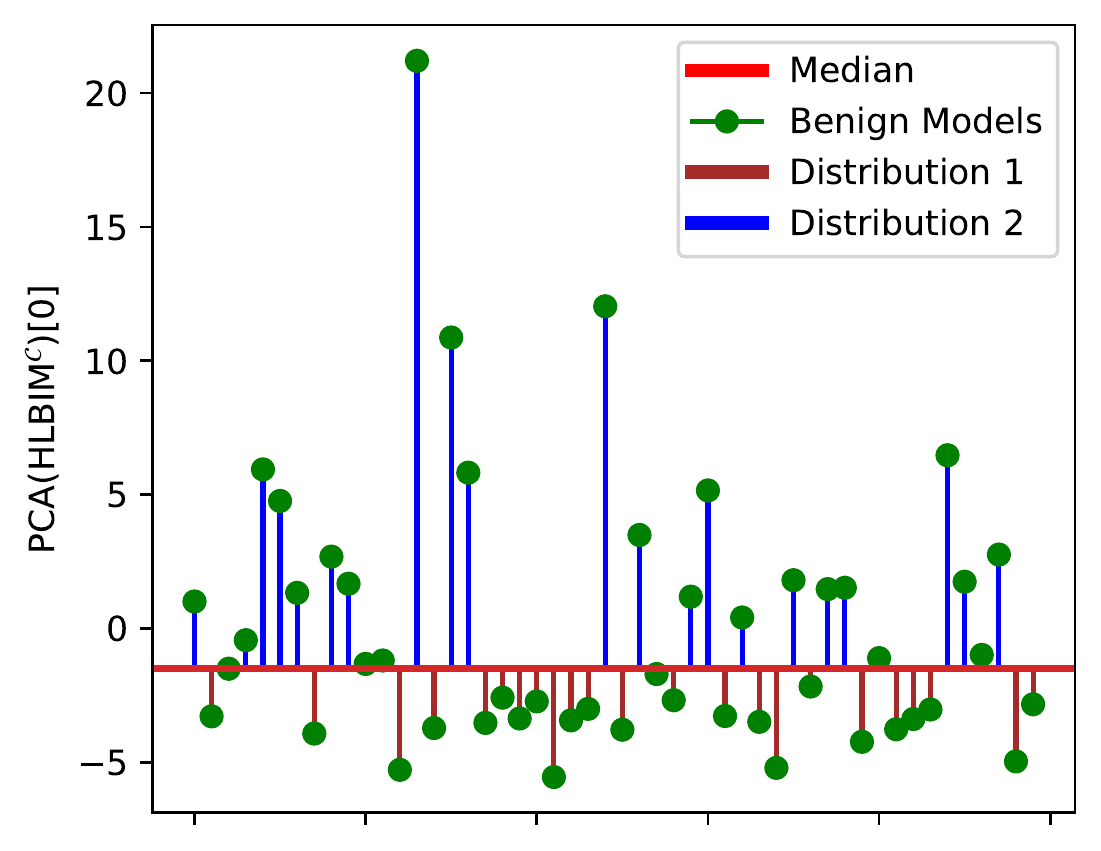} }}%
    \qquad
    \caption{Visualized distributions generated by \hyperref[alg:significance]{\alg\ref{alg:significance}} for pruning \ourname with $n = 100$, $PMR = 40\%$. Distributions are considered to differ significantly, indicating backdoors found from (a) - (c). Poisoned models are pruned iteratively.}%
    \label{fig:significance}
    
\end{figure*}

\begin{algorithm}[tb]
	\caption{\scalebox{0.76}{Voting Decision via Model Pruning for validation client $V_j$}}
	\label{alg:prune} 
	\scalebox{0.75}{
	\begin{minipage}{1.33\columnwidth}
	\small
	{
	\begin{algorithmic}[1]

		\State \textbf{Input:}\\
		\clientMetricNoSpace$^{C/E}_{m_j\;m\;l}$ \Comment{\clientMetric matrices for Cosine \& Euclidean distances}
		
		\State \textbf{Output:}\\
		client\_voting \Comment{Binary vector with client decisions for each $L^{t}_i$}

		\State \Comment{Analyze \clientMetric via dimension reduction}
		\State client\_voting\_model\_is\_benign $\gets$ Array of $|i|$ ones
		\For{dist\_type$^{C/E}$ in [COSINE-distance; EUCLIDEAN-distance]}
		    \State significant $\gets$ True
    		\State pruned\_models $\gets$ \{\}
		    \While{significant}
		        \State \Comment{Filter already pruned models}
		        \State \mbox{\clientMetric\_pruned} $\gets$ {\clientMetricNoSpace$^{C/E}_{m_j\;m\;l}$} $\forall$ $m \notin$ pruned\_models 
		        \State \Comment{Analyze remaining models}
    		    \State pc\_dim1\_values $\gets$ PCA(\clientMetric\_pruned)[0]
    		    \State significant $\gets$ SIGNIFICANCE(pc\_dim1\_values)
    		    \State malicious\_models $\gets$ \{\}
    		    \If{significant}
    		    \State clusters $\gets$ AGGLOM(nclusters = 2, pc\_dim1\_values)
    		    \State malicious\_models $\gets$ MIN\_CLUSTER(clusters).models()
    		    \EndIf
    		    \State pruned\_models.add(malicious\_models)
    		\State \Comment{Safety Abort Criterion}
                \If{$|$pruned\_models$|$ $>$ FLOOR($|L^{t}_i|$ -1) / 2}
                    \State malicious\_models.remove(MIN(malicious\_models))
    		      \State significant $\gets$ False
    		    \EndIf
          
    		    \State client\_voting\_model\_is\_benign[malicious\_models.indices] = 0
    		\EndWhile
		\EndFor
	\end{algorithmic}}
\end{minipage}}

\end{algorithm}
The PCA reduces the two-dimensional matrix (models $m$ $\times$ layers $l$) to a single dimension by analyzing the Principal Component (PC) values of the first dimension (pc\_dim1\_values in \hyperref[alg:prune]{\alg\ref{alg:prune}}). 
To detect the presence of backdoored models, we rely on statistical significance tests on the first PCA dimension.\footnote{\hyperref[app:algSignificance]{\app\ref{app:algSignificance}} discusses, that using multiple PC dimensions is suboptimal.} The intuition is that the median PC value is benign (cf. \hyperref[sec:problem-advmodel]{\sect\ref{sec:problem-advmodel}}). If all models are benign, the PC values follow a similar distribution and with this, the absolute distances to the median value of the PC values \textit{above} and \textit{below} the median value should follow the same distributions. A visualization of such two distributions is shown in \hyperref[fig:significance]{\fig\ref{fig:significance}}, where the blue and dark red lines represent the two mentioned distributions.

To compare and identify significant differences between these distributions, we analyze their \textit{means}, \textit{variances}, and \textit{outliers} leveraging different significance tests  (\mbox{line 15} in Alg.~\ref{alg:prune}). Outliers are considered since in general the mean and the variance of a distribution is not necessarily affected by few outlier samples.
First, \ourname forces an equal mean via \mbox{Student-T-Test}~\cite{ttest}. However, the variance can differ significantly even for equal means, allowing an adversary to adapt the backdoors according to these metrics. Therefore, \ourname checks for matching variance via \mbox{F-Test}\footnote{Levene-Test~\cite{ftest_levene} to asses equal variances.}.

To enhance robustness and prevent the adversary from attempting to fool the \mbox{F-Test}, we also employ the D-Test (Kolmogorov-Smirnov) to analyze the overall distributions. This additional test ensures equal goodness of fit of the distributions. We set a significance level of 0.01 for each test. \footnote{Typically, statistical tests use a significance level of 0.05. By using a lower p-value, we aim to reduce the likelihood of False-Positives and increase the sensitivity of \ourname.} Upon passing these tests, we investigate outliers that might not influenced the former metrics. These outliers represent weakly hidden poisoned models that would have a high impact on the aggregated model. To identify such outliers, we set two thresholds: 1)~We analyze the interquartile range of all data points by using a boxplot. 2) We analyze the distance of each point regarding the interval spanned by the 3$\sigma$-rule~\cite{pukelsheim1994three}. Data points lying outside the interval are marked as significant.\footnote{The boxplot focuses on detecting single outliers not violating the 3$\sigma$-rule~\cite{pukelsheim1994three}, while the 3$\sigma$-rule outperforms the boxplot for multiple outliers whose distance to the mean grow increasingly.} The algorithm is provided in \hyperref[app:algSignificance]{\app\ref{app:algSignificance}}.

If the tests indicate the presence of poisoned models, we employ hierarchical agglomerative clustering~\cite{frank16agglomerative} to generate two clusters and prune the models located in the smaller one (\mbox{lines 16-21})\footnote{Notably, the median is not suitable to separate both groups as it would always split the data points into two equally-sized groups. However, if one of them is discarded but significantly less than 50\% of models are poisoned, this will result in many FPs.}. This pruning is repeated until the significance tests report the absence of suspicious models (line 15)\footnote{Additionally, as a fine-tuning step, we add an abort criterion to stop the pruning process if more than \nicefrac{n}{2} -1 models have been pruned. This criterion is included to prevent obvious False Positives in a small number of experiments. While this may theoretically allow for False Negatives, i.e., undetected poisoned models due to the abort criterion coinciding with False Positives, our experiments (see \sect\ref{sec:eval}) show that the backdoor leads to significant differences within certain DLOs even for benign input samples. Consequently, the PC values of poisoned models differ more from the median (benign) value compared to benign models, ensuring that malicious models are consistently identified first.}. Due to the iterative pruning approach, we can detect and remove different backdoors, within one FL round $t$, that would not be detectable all together in a single cluster. The iterative pruning approach enables us to detect and remove different backdoors within a single federated learning round, which may not be detectable as a whole in a single cluster.

One pruning sequence is visualized in \hyperref[fig:significance]{\fig\ref{fig:significance:1}}-\hyperref[fig:significance]{\fig\ref{fig:significance:3}}. In \hyperref[fig:significance]{\fig\ref{fig:significance:3}}, no more significant models are detected. Another example can be seen in \hyperref[fig:graphs:pca1]{\fig\ref{fig:graphs:pca1}} to \hyperref[fig:graphs:pca3]{\fig\ref{fig:graphs:pca3}}. 

\begin{algorithm}[b]
	\caption{Voting Filtering via Stacked Clustering}
	\label{alg:stackedclustering} 
	\scalebox{0.75}{
	\begin{minipage}{1.33\columnwidth}
	\small
	{
	\begin{algorithmic}[1]
			
			\State \textbf{Input:}\\
			voting\_matrix \Comment{Matrix of client votes. Dimensions: ($|C_i| \times |L^{t}_i|$)}
			\State \textbf{Output:}\\
			aggregated\_voting \Comment{List with final voting decisions for each $L^{t}_i$}
			\State \Comment{Majority Cluster Detection}
			\State clusters $\gets$ AGGLOM(nclusters = 2, voting\_matrix)
			\State majority\_cluster $\gets$ MAX\_CLUSTER(clusters) 
			
			\State \Comment{Miss-Classification Compensation}
			\State filtered\_cluster $\gets$ DBSCAN(majority\_cluster, min\_samples=1, $\epsilon$=0.5)
			\State aggregated\_voting $\gets$ \{"\textit{all-benign}"\}
			\State max\_count $\gets$ 0
			\For{cluster in filtered\_cluster}
				\State count $\gets$ $|$cluster$|$
				\If{count $>$ max\_count}
				\State aggregated\_voting $\gets$ cluster[0].decision
				\State max\_count $\gets$ count
				\EndIf
			\EndFor
	\end{algorithmic}}
\end{minipage}}

\end{algorithm}
\subsection{Voting Aggregation}
\label{sec:approach-voting}
\noindent \textbf{Stacked Clustering:}
After the clients provided their votes in the form of binary vectors to the server\footnote{It is worth noting, that each client does not evaluate its own local model, but just reports it as benign by default.}, the server is confronted with potentially malicious votes from adversaries as well as with unintentional wrong votes of benign clients, which can occur if a model exceeds, i.e., an outlier threshold slightly.
To address this issue, we employ a two-level stacked clustering, that selects the most representative voting vector from all submissions. The purpose of the first level is to eliminate obvious malicious votes by pruning the smaller of two clusters, as due to the majority assumption the larger cluster has to be the benign one. The second clustering is a \mbox{plain-majority-voting} on the rest, which are expected to be mostly benign votes. Thus, this step ensures the robustness of the aggregation against minor misclassifications of benign clients and adversaries, deviating only slightly from benign votes to remain inconspicuous.

In comparison, for plain majority voting, in a scenario with 49\% adversarial clients who vote for all malicious models to be benign, a single incorrect vote of one benign client for a poisoned model can result in the acceptance of this model. 

As depicted in \hyperref[alg:stackedclustering]{\alg\ref{alg:stackedclustering}} in detail, the server first generates two clusters on the binary voting vectors by agglomerative clustering~\cite{frank16agglomerative} and identifies the bigger one as votes from benign clients, which is reasonable due to the majority assumption (cf. \hyperref[sec:problem-advmodel]{\sect\ref{sec:problem-advmodel}}). However, this cluster can contain minor errors of benign clients as well as malicious clients that manipulate their voting to be similar to benign behavior. Therefore, we conduct a second clustering to extract the most frequent binary voting vector by using DBSCAN~\cite{ester1996density} and inspecting the cluster sizes of the output. The voting of the biggest cluster is the final result of the voting aggregation.

\noindent\textbf{Robustness:} The stacked clustering ensures robustness in scenarios where every malicious client marks every benign model as malicious and vice versa, since these manipulated votes are removed after the first clustering.
Malicious feedback, where malicious clients vote as benign as possible, trying to invert the decision for one specific model, will be mitigated by the second clustering. The same holds for minor voting errors of benign clients. Leveraging the stacked clustering approach, which relies on the majority assumption, the algorithm remains robust against PMRs of up to 49\%.

This strategy of first identifying benign votes by majority and then selecting the best voting via majority as the final decision outperforms na\"ive majority voting, which would not ignore False-Positives of benign clients, making it less robust. We discuss respective experiments in \hyperref[sect:eval-clustering]{\app\ref{sect:eval-clustering}}.

\section{Evaluation}
\label{sec:eval}
\noindent In this section, we first depict our experimental setup in \hyperref[eval:setup]{\sect\ref{eval:setup}} and then describe the influence of various parameters in \hyperref[eval:params]{\sect\ref{eval:params}}. Afterward, in \hyperref[sec:eval-perf]{\sect\ref{sec:eval-perf}}, we investigate the runtime performance of our approach.

\subsection{Experimental Setup}
\label{eval:setup}
\noindent To simplify the comparison of our evaluation with other poisoning defenses, we aligned our experimental setup with recent works~\cite{bagdasaryan,cao2021fltrust,rieger2022deepsight}, as we describe in the following.

\begin{figure*}[thb]
    \centering
    \subfloat[\label{fig:graphs:global}\centering Plain Cosine Distance]{{\includegraphics[height=2.4cm,trim={0 0.5cm 0 0}]{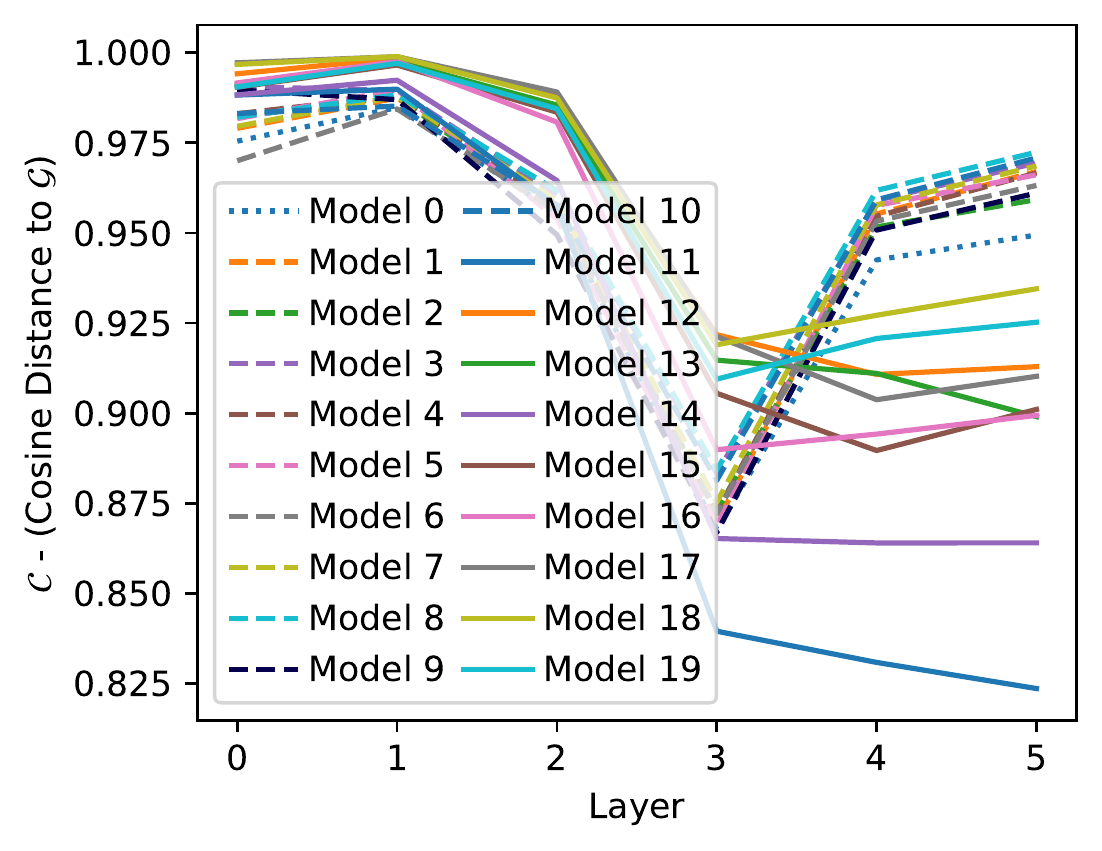} }}%
    \qquad
    \subfloat[\label{fig:graphs:squared}\centering \clientMetricNoSpace$^C$ for one label]{{\includegraphics[height=2.4cm,trim={0 0.5cm 0 0}]{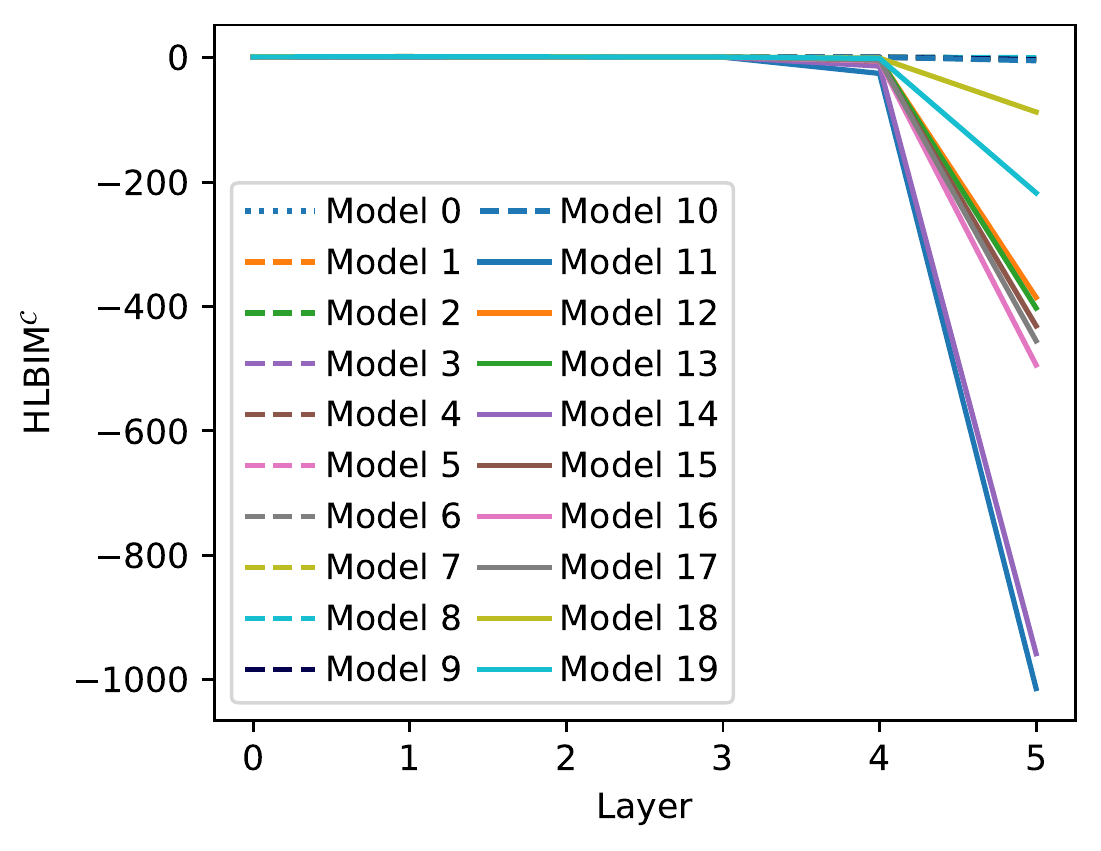} }}%
    \qquad
    \subfloat[\label{fig:graphs:concat}\centering \clientMetricNoSpace$^C$]{{\includegraphics[height=2.4cm,trim={0 0.5cm 0 0}]{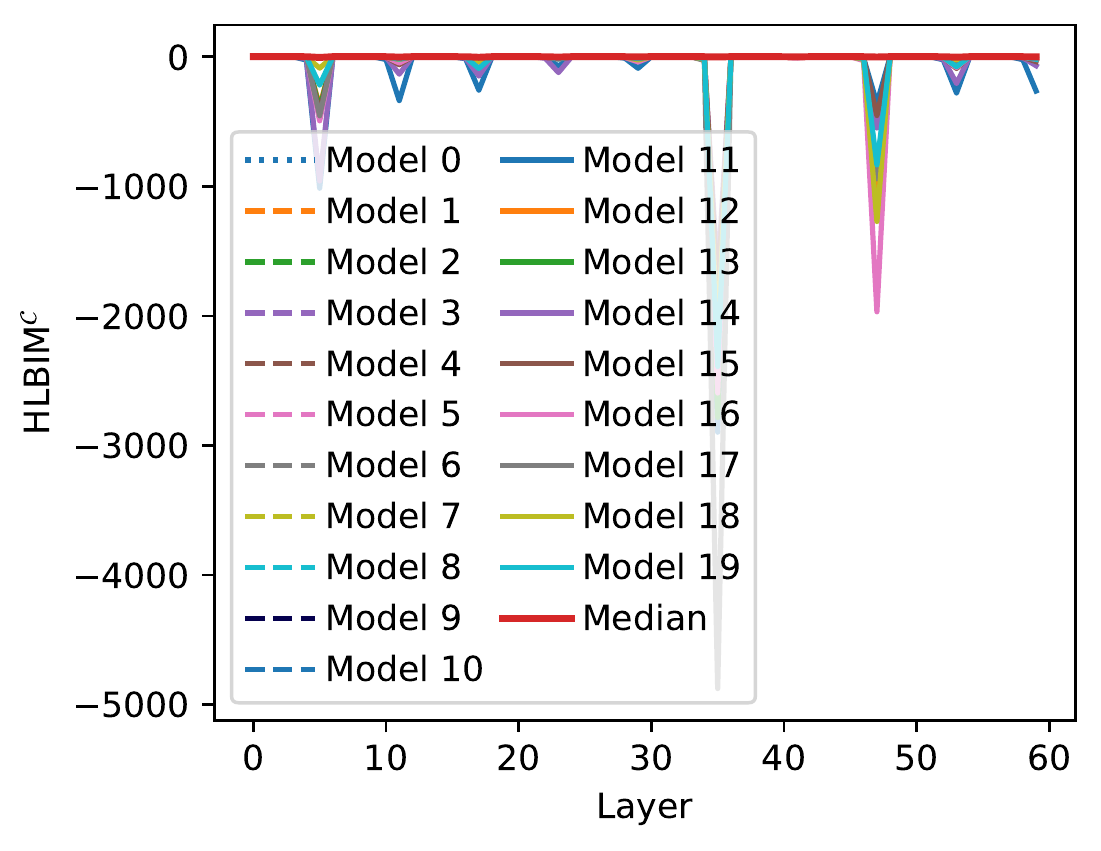} }}%
    \qquad
    \subfloat[\label{fig:graphs:concatclean}\centering \clientMetricNoSpace$^C$ for benign models only]{{\includegraphics[height=2.4cm,trim={0 0.5cm 0 0}]{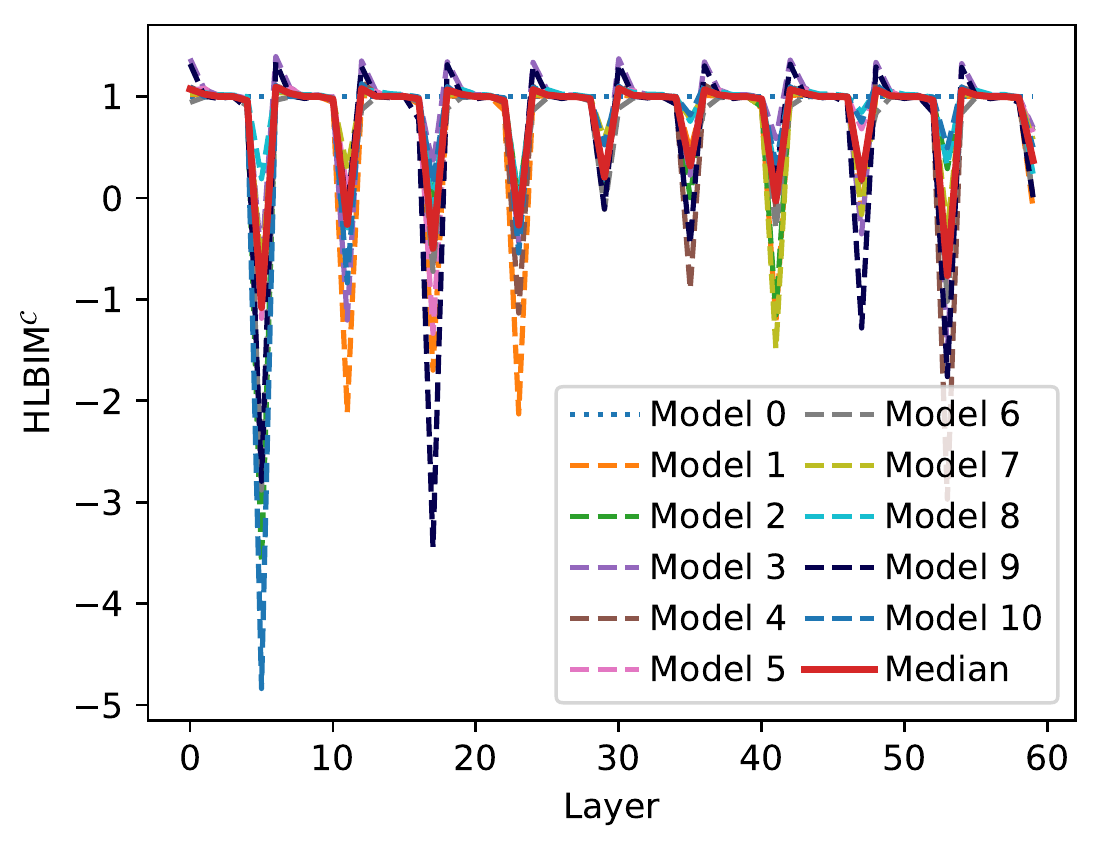} }}%
    \qquad
    \subfloat[\label{fig:graphs:pca1}\centering PCA in round 1]{{\includegraphics[height=2.4cm,trim={0 0.5cm 0 0}]{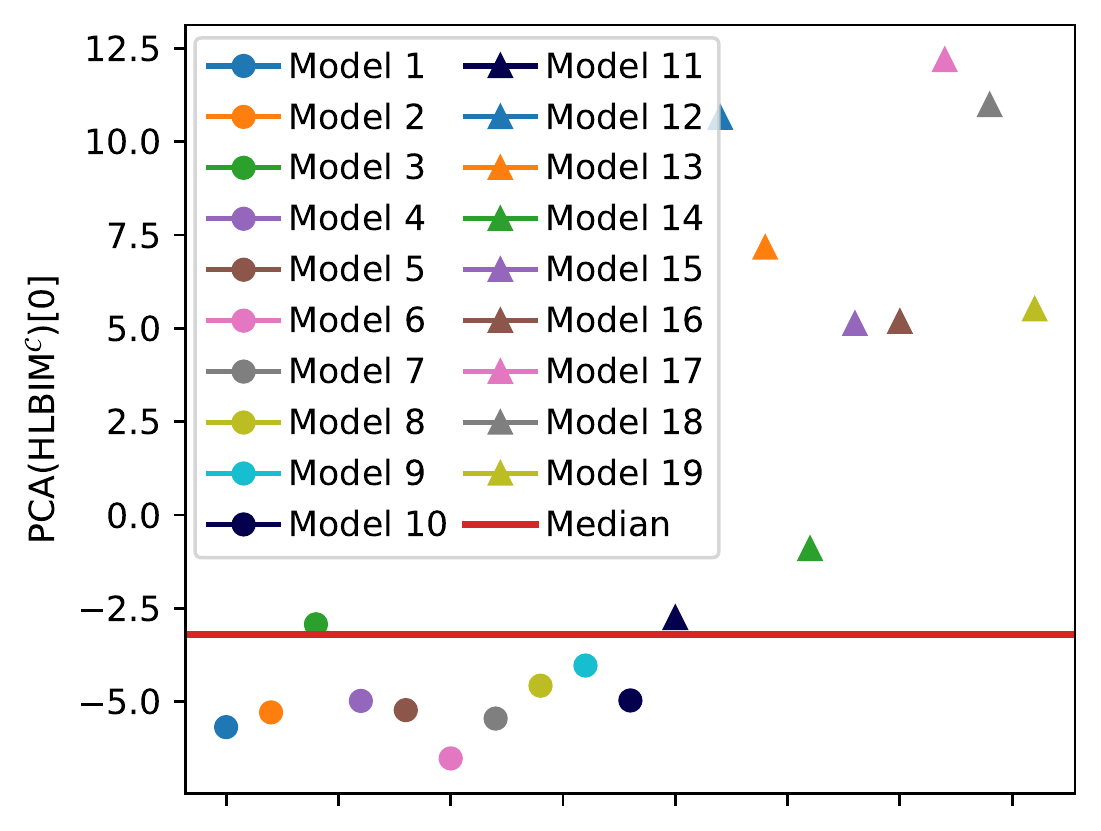} }}%
    \qquad
    \subfloat[\label{fig:graphs:pca2}\centering PCA pruning round 2]{{\includegraphics[height=2.4cm,trim={0 0.5cm 0 0}]{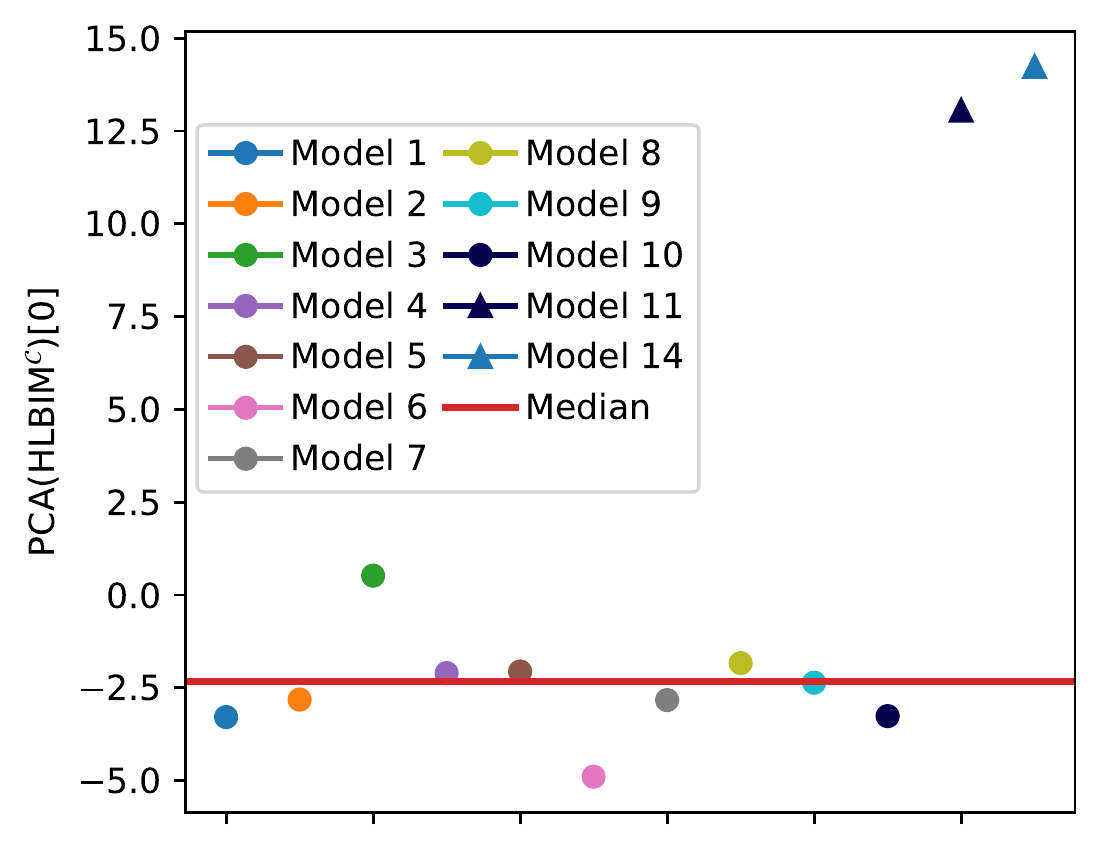} }}%
    \qquad
    \subfloat[\label{fig:graphs:pca3}\centering PCA pruning round 3]{{\includegraphics[height=2.4cm,trim={0 0.5cm 0 0}]{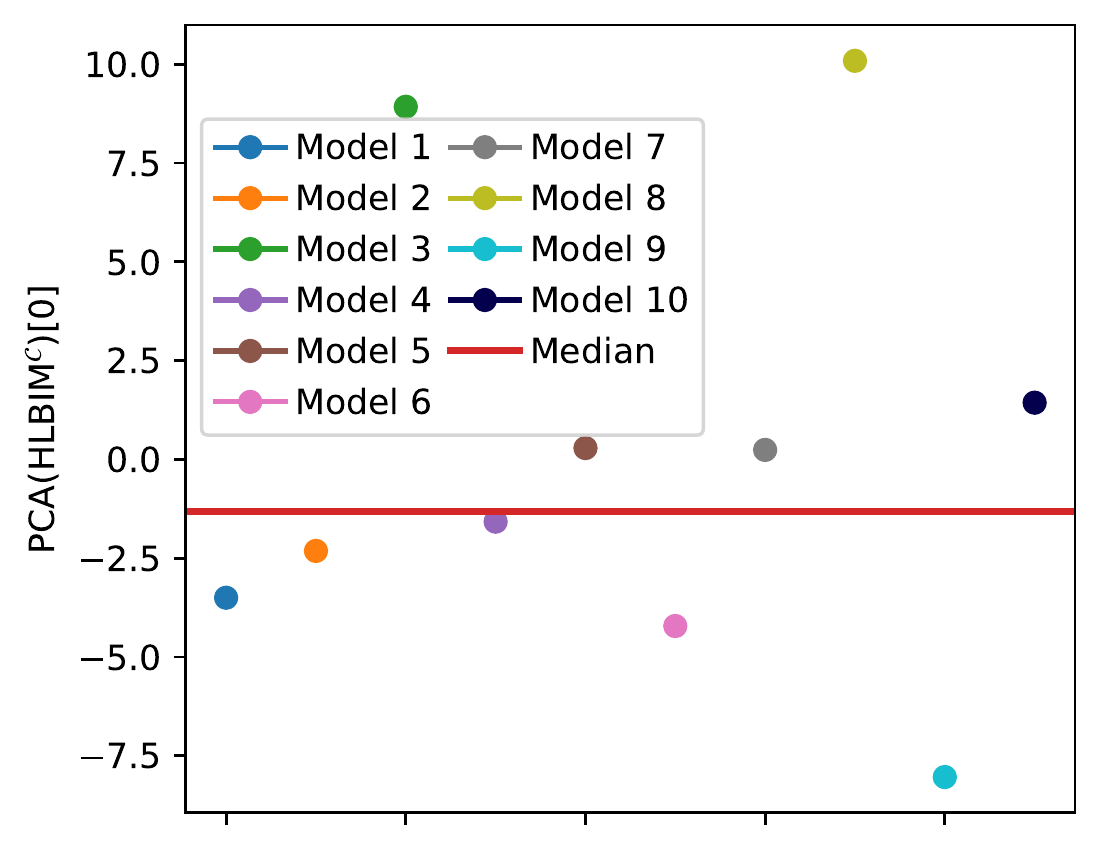} }}%
    \qquad
    \subfloat[\hspace{-0.035cm}\label{fig:graphs:outlier}\centering \hspace{-0.035cm}Outlier detection boxplot]{{\includegraphics[height=2.4cm,trim={0 0.5cm 0 0}]{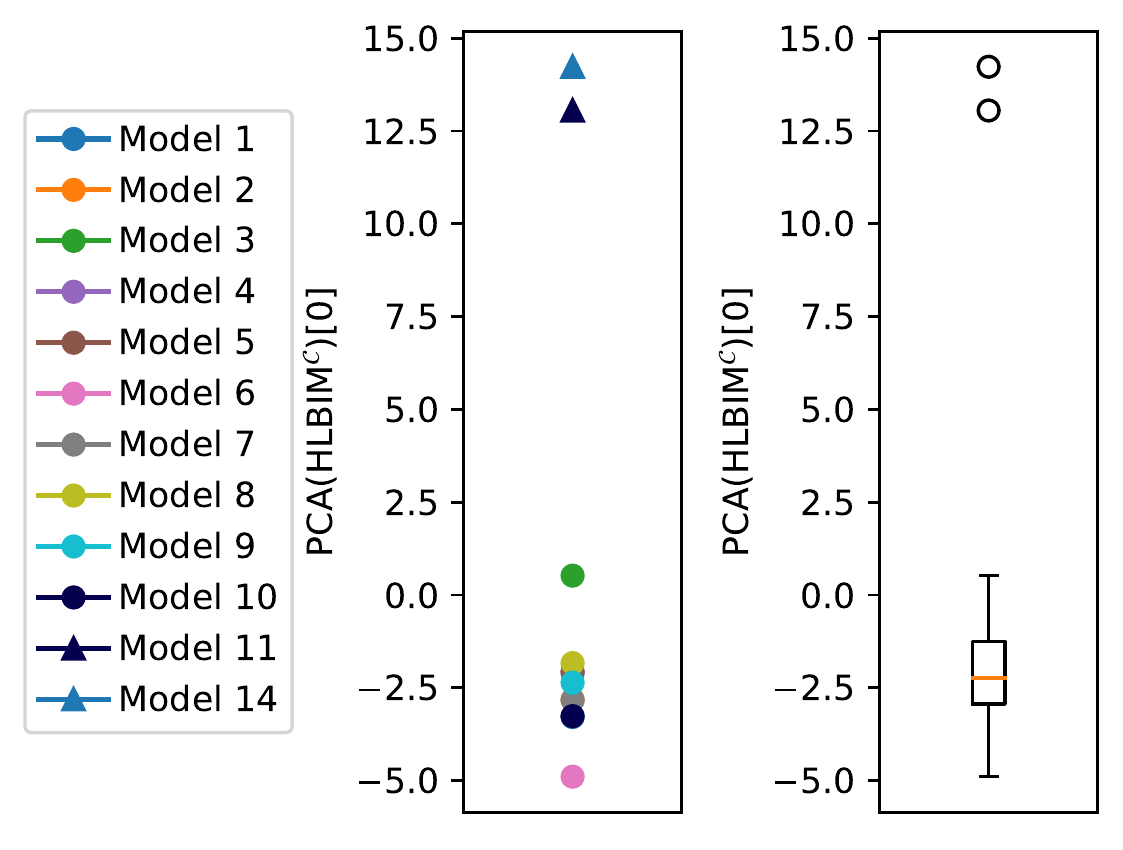} }}%
    \caption{Visualization of intermediate outputs of \ourname with default configurations (models 0 to 10 are benign).}
    \label{fig:graphs}%
\end{figure*}
\noindent\textbf{Computational Setup:}
All experiments were implemented in Python using the Deep Learning library PyTorch~\cite{pytorch}. The experiments were executed on a server with an Intel Xeon 5318S with Intel SGXv2, 2 Nvidia RTX A6000, and 512 GB main memory, from which 128GB were reserved as secure memory. For executing Python code inside an SGX enclave, we leveraged the library Gramine~\cite{tsai2017graphene}.

\noindent\textbf{Datasets:}
For our evaluation, we use the popular benchmark datasets CIFAR-10~\cite{krizhevsky2009learning}, consisting of 50k training images and 10k test images, and the MNIST~\cite{deng2012mnist} dataset, consisting of 60k training images and 10k test images. Both datasets contain samples from 10 classes and are frequently used for evaluating poisoning defenses~\cite{bagdasaryan,cao2021provably,fung2020FoolsGold,mcmahan2017,munoz19AFA,nguyen22Flame,rieger2022deepsight}. To simulate the FL setup, we split the training dataset into local datasets $D_k \in \{D_1, \ldots, D_\nClients\}$ consisting of \numprint{2560} samples\footnote{{The dataset size of 2560 samples is chosen based on other approaches (between 600~\cite{bagdasaryan, zhao2020shielding} and 2000~\cite{shen16Auror}) to ease the comparison.}}, one $D_k$ for each $C_k \in \{C_1, \ldots, C_\nClients\}$. Aligned with existing work~\cite{cao2021fltrust,rieger2022deepsight}, we created datasets for different \nonIid scenarios:~1) For \mbox{1-class} \nonIid with a \nonIid rate $q$, a main label is chosen randomly for each client and $q$ percent of samples in $D_i$ are changed to samples with the respective main label, while the remaining labels are chosen from all labels uniformly distributed. Therefore, for a \nonIid rate of $q=1.0$, each client only uses data from its main label(s), such that the data of different clients are disjoint if they have different main labels. 2) For \mbox{2-class} \nonIid we do the same but choose the subsequent label of the main label as "second main label". 3) For Dirichlet and Normal distribution, we produced label counts according to the respective distribution.

For the \cifar dataset, we used a light version of the Resnet-18 network, as described by Bagdasaryan \etal~\cite{bagdasaryan}, which delivers five deep layers and one final layer output. For MNIST, we reimplemented a version of the Convolutional Neural Network (CNN), following Cao \etal~\cite{cao2021provably}.

\noindent\textbf{Default Configurations:}
\label{eval:setup:config}
The default parameters and configurations, including hyperparameters for our experiments, are provided in \hyperref[eval:params]{\tab\ref{tab:defaultconf}} in \hyperref[app:experimentparams]{\app\ref{app:experimentparams}}. In our default setup, we use the CIFAR-10~\cite{krizhevsky2009learning} dataset, an adaptive adversary leveraging a \constrainandscale attack~\cite{bagdasaryan} to inject a semantic backdoor, a Poisoned Data Rate (PDR) of 0.1, a PMR or 0.45 and set $\alpha$ to 0.7. We utilize $|C_i|=20$ clients to participate in the FL round $t$ as well as in the feedback loop. We select the main label of each client according to its index $i$, so that we have the most disjoint label settings and prevent getting multiple clients with the same main label by chance.

\subsection{Outputs and Influence Factors}
\label{eval:params}
\noindent In \hyperref[sec:eval-params-output]{\sect\ref{sec:eval-params-output}}, we visualize and explain the output of our experiments and list all configurations. Afterward, we discuss the parameters and other influencing factors in \hyperref[sec:eval-params-influence]{\sect\ref{sec:eval-params-influence}.}
\subsubsection{Experiment Outputs}
\label{sec:eval-params-output}
To improve the comprehension of our approach, in \hyperref[fig:graphs]{\fig\ref{fig:graphs}} we visualize the intermediate outputs of our algorithm. The experiment is conducted with our default configurations from \hyperref[tab:defaultconf]{\tab\ref{tab:defaultconf}} and therefore contains 20 clients. \hyperref[fig:graphs]{\fig\ref{fig:graphs:global}} depicts the plain Cosine distance from each local model $L^{t}_i$ to the global model $G^t$ for one label (in this case, label 8) averaged over that label. As explained in \hyperref[sec:approach-validationAlgorithm]{\sect\ref{sec:approach-validationAlgorithm}}, from this metric alone, one cannot clearly identify the poisoned models.
To enable backdoor detection, we produce the matrices containing our novel metric \clientMetric, which can be seen for Cosine distance in \hyperref[fig:graphs]{\fig\ref{fig:graphs:squared}} for one label. Subsequently, in \hyperref[fig:graphs]{\fig\ref{fig:graphs:concat}}, we can then observe the whole \clientMetricNoSpace$^{C}$ plot. The malicious models are responsible for the peaks. This can be analyzed by comparing the values for \clientMetricNoSpace$^{C}$ to the ones in the poisoned-free version in \hyperref[fig:graphs]{\fig\ref{fig:graphs:concatclean}}. According to \hyperref[alg:prune]{\alg\ref{alg:prune}}, we conduct the PCA on the \clientMetric matrices to obtain \hyperref[fig:graphs]{\fig\ref{fig:graphs:pca1}}, which is used in the first pruning round of our significance test. The results of the second and third rounds of pruning are depicted in \hyperref[fig:graphs]{\fig\ref{fig:graphs:pca2}} and \hyperref[fig:graphs]{\fig\ref{fig:graphs:pca3}}. At this step, we end up with just benign models, which results in negative significance tests. \hyperref[fig:graphs]{\fig\ref{fig:graphs:concatclean}} shows the cleaned \clientMetricNoSpace$^{C}$ graph used to produce the last PC values in \hyperref[fig:graphs]{\fig\ref{fig:graphs:pca3}}. \hyperref[fig:graphs]{\fig\ref{fig:graphs:outlier}} shows an exemplary boxplot of our outlier detection algorithm during the second pruning round (cf. \hyperref[fig:graphs]{\fig\ref{fig:graphs:pca2}}). The abnormalities in the malicious models derive from their local BA, which in our experiments is always 100\%. This means the attacker is able to incorporate the backdoor in his local model.

\noindent\textbf{Conducted Experiments:}
\hyperref[tab:experiments]{\tab\ref{tab:experiments}} lists all of our conducted experiments, each changing one parameter of the default configuration. The used metrics are defined in \hyperref[tab:experiments]{\app\ref{app:metrics}}. Additionally, we tested the defense against two untargeted poisoning attacks: 1) We randomly selected the labels for each sample in the training and test set. 2)~We changed the learning algorithm to maximize the loss function. Both experiments delivered 100\% defense success rate. Thus we can claim that \ourname also reduces the risk of \textit{untargeted poisoning attacks}. Furthermore, the following two combined attacks that integrate two backdoors with different triggers and target labels $T_\adversary$ at once have been evaluated and results are reported in \hyperref[tab:experiments]{\tab\ref{tab:experiments}}. The following two scenarios were considered: 1)~Every \adversary attempts to inject both backdoors. 2)~Half of the malicious clients integrate one backdoor, and the other half integrate a different backdoor.

\subsubsection{Influence Factors}
\label{sec:eval-params-influence}
In \hyperref[app:parameter-influence]{\app\ref{app:parameter-influence}} we evaluate different factors that might affect the performance of \ourname ($\alpha$, the PDR, and the distribution of local data). However, in all cases \ourname effectively identifies benign and poisoned models independently from the scenario. Thus, it achieves 100\% True-Positive-Rate (TPR) as well as True-Negative-Rate (TNR). Further, we observe that \ourname has no negative impact on the MA (fulfilling \requirement{2}). The same perfect detection rates are achieved in our other experiments, listed in \hyperref[tab:experiments]{\tab\ref{tab:experiments}}.

\begin{tablefloat*}
\begin{center}
\scaleTable{
\begin{tcolorbox}[tab3,tabularx={c || Z | c | c | c | c},boxrule=0.75pt]
\textbf{Analyzed Parameter} & \textbf{Parameter Values} & \textbf{TPRs} & \textbf{TNRs} & \textbf{FPRs} & \textbf{FNRs}\\\hline\hline
Data distributions &  \makecell{CIFAR-10, \mbox{1-class} \nonIid, $q = [0.0, 0.1 ,... ,1.0]$ \\ CIFAR-10, \mbox{2-class} \nonIid, $q = [0.0, 0.1 ,... ,1.0] $ \\ CIFAR-10, Dirichlet\\CIFAR-10, Normal\\MNIST \mbox{1-class} \nonIid, $q = [0.0, 1.0]$} & \makecell{100\% \\ 100\%\\ 100\% \\ 100\%\\ 100\%} & \makecell{100\% \\ 100\%\\100\% \\ 100\%\\ 100\%}& \makecell{0\% \\ 0\%\\ 0\% \\ 0\%\\ 0\%} & \makecell{0\% \\ 0\%\\0\% \\ 0\%\\ 0\%}\\
 \hline
Adversarial adaptation rate $\alpha$ & $\alpha = [0.1, 0.2 ,... ,0.9]$ & 100\% & 100\% & 0\% & 0\%\\
\hline
Poison Data Rate (PDR) & $pdr = [0.1, 0.2 ,... ,0.9]$ & 100\% & 100\% & 0\% & 0\%\\
\hline
Poison Model Rate (PMR) \& number of clients $n$ & \makecell{$pmr(n=20) = [0.05, 0.1 ,... ,0.45]$\\$pmr(n=100) = [0.01, 0.2 ,... ,0.49]$\\$pmr(n=10) = [0.1, 0.2,... ,0.4]$, (\mbox{1-class} \nonIid $q=0.0$)} & \makecell{100\% \\ 100\%\\ 100\%} & \makecell{100\% \\ 100\%\\100\%} & \makecell{0\% \\ 0\%\\ 0\%} & \makecell{0\% \\ 0\%\\0\%}\\
\hline
Poisoning &  \makecell{Pixel Backdoors , Label Swap, Semantic \\ 2 combined attacks\\ 2 untargeted attacks} & \makecell{100\%  \\ 100\% \\ 100\%} & \makecell{100\% \\ 100\%\\100\%} & \makecell{0\%  \\ 0\% \\ 0\%} & \makecell{0\% \\ 0\%\\0\%}\\
\hline
Starting FL round $t$ & \makecell{$t = 1000$\\$t = 0$\\$t >= 1$} & \makecell{100\% \\ 100\% \\ 100\% } & \makecell{100\% \\ 0\% \\ 100\%} & \makecell{0\% \\ 100\% \\ 0\% } & \makecell{0\% \\ 0\% \\ 0\%}\\
\hline
Malicious Training Learning Rate(M-LR) & \makecell{$LR = [0.01, 0.001]$\\} & \makecell{100\%} & \makecell{100\%} & \makecell{0\%} & \makecell{0\%}\\
\hline

\end{tcolorbox}}
\captionsetup{type=table}
\captionof{table}{Listing of conducted experiments with TPR, TNR, FPR, and FNR of \ourname. The default settings are used and only the analyzed parameter is changed for one experiment. Multiple parameters within brackets denote multiple experiments.}
\label{tab:experiments}
\end{center}
\end{tablefloat*}

\noindent\textbf{Disjoint Data Scenario:} We conducted an experiment with only six benign and four malicious clients in \mbox{1-class} \nonIid for $q=[0.0]$ and assigned samples from each of the ten existing labels in the dataset to a different client. This setup results in completely disjoint training data, which reflects a full \nonIid scenario. \ourname was also effective in the detection of the four malicious clients even in this \mbox{edge-case} scenario, fulfilling \challenge{1}.

\noindent Besides that, we tested a scenario, where all 20 clients were benign and changed this setting to a \mbox{1-class} \nonIid scenario, where 40\% of clients possess the same main label, e.g., we assigned only red cars from \cifar. We observed that the defense did not falsely filter out benign models and hence such a scenario does not negatively affect the convergence.

\noindent\textbf{MNIST:} In addition, we evaluated \ourname on the MNIST dataset in \mbox{1-class} and \mbox{2-class} \nonIid with $q=[0.0,1.0]$ scenarios starting from FL round 100 using the Label Swap backdoor and also for a smaller learning rate of 0.001 at the malicious clients. This showed that our approach is not restricted to one specific dataset, thus not hindering deployment in real-world scenarios.

\noindent\textbf{Varying PMR:} Our experiments are conducted with the biggest Poisoned Model Rate (PMR) possible regarding the $C_i$ clients and, since we are pruning the malicious models, a smaller PMR would still result in the same outcome. Hence, we can conclude that the PMR has no influence. In the default setting, the PMR has a maximum value of $\nicefrac{9}{11} = 45\%$. To stress this parameter to the maximum, we conducted an additional experiment with $n=100$ clients and a PMR of 49\%. The results also show perfect detection rates of 100\%, making the approach \mbox{effective in small and large FL setups.}

\noindent\textbf{Randomly Initialised Model:} The only exception with regard to \ournameGen effectiveness occurred in the first round ($t=0$) when starting with a randomly initialized model. Here, \ourname accepted all models, including the poisoned models. However, as the model's parameters of all models were changed significantly, the impact of the poisoned models was negligible. Therefore, the BA remained 0\% and the attack was not effective, while the adversary was not able to inject the backdoor in later rounds ($t\geq1$). This experiment showed two facts: 1) It is harder for an attacker to implement the backdoor as long as the model did not converge to a certain MA, as already discussed in earlier work~\cite{bagdasaryan}. 2) Nevertheless, \ourname already detected 100\% of malicious models in round $t=1$, meaning our approach is not limited to already converged global models $G^t$.

\noindent\textbf{Robustness against Adaptive Adversaries:}
An adversary can adapt to the defense in various ways by integrating a second loss function (cf.~Bagdasaryan \etal~\cite{bagdasaryan}). Thereby, the adversary can try to minimize the distance of the model weights to the global model $G^t$ or first train a benign local model and then try to adapt precisely to our defense algorithm by leveraging DLOs measured on that model. We conducted experiments with both adaptation strategies, finding that the former delivered better results for the adversary, meaning that poisoned models were harder to identify by \ourname (regarding the significance level). Thus the more difficult-to-handle adaptation method is part of our default setting but does not prevent \ourname from detecting the backdoor. The most relevant reason for that is, that the adversary cannot adapt to the other clients' local data. If the adversary increases the level of adaptiveness extremely, he fails in introducing the required BA in the local model, so that his contribution is averaged out, even without clipping or noising methods (cf. adversarial's dilemma in \hyperref[sec:background:bdfl]{\sect\ref{sec:background:bdfl}}), fulfilling \requirement{1}.

As the client-side validation algorithm compares the PC values of the individual models against each other, a sophisticated adversary might try to adapt its attack by splitting all malicious clients into different groups leveraging different PDRs. To show \ournameGen robustness in such scenarios, we conducted an experiment where two groups of malicious clients use different PDRs of 10\% and 30\%. However, the iterative running enabled \ourname to identify all benign and poisoned models effectively.

Overall, \ourname is robust in various scenarios, independent of specific FL system settings, and therefore applicable in real-world scenarios. The reasons for the high success rates are that the usage of benign validation data allows a detailed analysis of the local models' behavior. In addition, benign clients rarely deliver wrong votes due to our significance test (cf. \hyperref[alg:significance]{\alg\ref{alg:significance}}) based on \clientMetric. Even if the malicious clients manipulate the voting of its secure enclave, they are compensated by the subsequent stacked clustering voting aggregation. 

\subsection{Runtime Overhead of \ourname}
\label{sec:eval-perf}

\noindent To analyze the performance overhead introduced by \ourname, we measured the runtime of the different phases of \ournameGen client-validation in a TEE (SGX) and compared it to the execution outside of a TEE on the CPU for our default setting with 20 models.
Further, we also measured the runtime when using the GPU for the predictions. The results, being averaged over ten executions, are shown in \hyperref[tab:evaltimes]{\tab\ref{tab:evaltimes}}. As the table shows, the overhead for the attestation is with 0.1s in average negligible, as well as the time for the pruning. The \clientMetric calculation takes similar time for all three versions, as it is done for all three platforms on the CPU, also if an accelerator is available. The main difference between the individual versions is the time that is needed for predicting the DLOs. Here, using an accelerator, i.e., a GPU, shows a significant performance improvement. We elaborate on the runtime overhead in \sect\ref{sec:discussion-overhead}.

\subsection{Comparison with Existing Approaches}
\noindent \hyperref[tab:successattacks]{\tab\ref{tab:successattacks}} shows the comparison of \ourname with several \sota IR approaches~\cite{fung2020FoolsGold,naseri2022local,zhao2020shielding}, as well as DF methods~\cite{blanchard17Krum,shen16Auror,yin2018Median} in our default scenario. 
Notably, existing approaches make certain assumptions about the attack strategy and data scenario: Zhao \etal assume that the attack reduces the MA~\cite{zhao2020shielding}, which does not always hold in practice (cf. O2) and hence, hinders detection. Median~\cite{yin2018Median}, instead, effectively mitigates the attack but cannot handle the \nonIid scenario and drops the MA to the performance of a na\"ive classifier. Existing DF approaches make assumptions about the attack strategy and data distribution: Auror~\cite{shen16Auror} clusters the parameters of all model updates into two clusters once and considers the smaller cluster as suspicious. This approach fails for the highly \nonIid scenario that we described in \hyperref[sec:background]{\sect\ref{sec:background}}, showing the advantage of \ournameGen iterative pruning. Krum assumes the benign models to have low distances among each other~\cite{blanchard17Krum}, which does not hold for the \nonIid scenario and can be easily circumvented by an adaptive adversary (cf.~\hyperref[sec:eval-params-influence]{\sect\ref{sec:eval-params-influence}}), while FoolsGold~\cite{fung2020FoolsGold} cannot handle the case that some benign clients have similar data, thus having updates that point in the same direction. In comparison, \ourname does not make any assumptions about the data scenario or the attack strategy while still being able to identify all poisoned and benign models correctly.

\begin{tablefloat}[b]
\begin{center}
\scaleTable{
\begin{tcolorbox}[tab4,tabularx={ l|| rrrrr|r },boxrule=0.75pt]
\textbf{Platform} & \textbf{Attestation} & \begin{tabular}[c]{@{}c@{}} \textbf{Model}\\      \textbf{Transmission}\end{tabular}  & \textbf{Predictions} & \textbf{\clientMetric}   & \textbf{Pruning} & \textbf{Total} \\\hline\hline
SGX & 0.1 & 4.0 &19.7 & 5.5 & 0.1 & 29.5\\
CPU & \multicolumn{1}{c}{-} &  2.6 & 10.5 & 5.0 & 0.0 & 18.1\\
GPU & \multicolumn{1}{c}{-} &  2.7&4.8 & 5.1 & 0.0 & 12.8\\

\end{tcolorbox}}
\captionsetup{type=table}
\captionof{table}{Average evaluation times of \ournameGen steps in SGX, outside SGX on a CPU and on a GPU in seconds.}
\label{tab:evaltimes}
\end{center}
\end{tablefloat}

It should be noted that \hyperref[tab:successattacks]{\tab\ref{tab:successattacks}} showing our default scenario contains \nonIid data distributions (disjoint data). As each client optimizes for its own training set, few benign models can negatively affect the MA on the test set containing all labels. In this special case the MA can improve by excluding benign models. For example, Krum selects one local model as new global model, which might have superior performance on the main task in the test set, but cannot prevent the backdoor from being active.

In a separate experiment, we also evaluated BaFFLe~\cite{andreina2020baffle} on \cifar data. Notably, BaFFLe first needs a benign warm-up phase without attacks~\cite{andreina2020baffle}\footnote{We discuss the real-world applicability of this precondition in \sect\ref{sec:sota}.}, which we set to 30 rounds starting from FL round 100. It is necessary to evaluate multiple rounds to analyze BaFFle's ability to accept a benign aggregation model. This ensures that BaFFle does not reject every update. Therefore, we performed this comparison in a separate experimental setting using a less converged model, hence these results are not included in \hyperref[tab:successattacks]{\tab\ref{tab:successattacks}}. We observed that BaFFLe was not resilient to the constrain-and-scale attack~\cite{bagdasaryan}, resulting in a BA of 80\%, while \ourname effectively identified also in this setting benign and poisoned models (TPR=100\%, TNR=100\%, BA=0\%).
\ourname is more efficient than BaFFLe, because \ourname analyzes the outputs of all layers, while BaFFLe focuses on the output of the last layer, which an adversary will always try to make inconspicuous (O2).

\subsection{Alternative Aggregation Techniques}
As described in \sect\ref{sec:background:fl}, we focus on a version of \fedavg for aggregation that weights all accepted clients' models equally, regardless of their dataset sizes. To show the general applicability of \ourname, we conducted an experiment where we weighted the accepted models' based on the dataset sizes which were reported by the clients. We observed the same performance as in our default setting (TPR=100\%, TNR=100\%) and the aggregated model achieved a MA of 62\%. Notably, the ability of \ourname to filter poisoned models is independent of the concrete aggregation function, as the aggregation is performed after the filtering process is finished.

\begin{tablefloat}[tb]%
\begin{center}
\scaleTable{
\begin{tcolorbox}[tab3,tabularx={l || rr | rrr},boxrule=0.75pt]
\textbf{Approach} & \textbf{BA} & \textbf{MA} & \textbf{TPR} & \textbf{TNR} & \textbf{PRC}\\\hline\hline
No Attack & 0.0 & 62.0 & & &\\
No Defense & 80.0 & 61.5 & & & \\\hline
Differential Privacy~\cite{naseri2022local} & 80.0 & 50.6 & \multicolumn{1}{c}{-} & \multicolumn{1}{c}{-} & \multicolumn{1}{c}{-}\\
Zhao~\etal~\cite{zhao2020shielding} & 100.0 & 61.2 & \multicolumn{1}{c}{-} & \multicolumn{1}{c}{-} & \multicolumn{1}{c}{-}\\
Median~\cite{yin2018Median} & 0.0 & 10.0 & \multicolumn{1}{c}{-} & \multicolumn{1}{c}{-} & \multicolumn{1}{c}{-}\\\hline
FoolsGold~\cite{fung2020FoolsGold} & 0.0 & 10.0 & 100.0 & 9.0 & 47.4\\
Krum~\cite{blanchard17Krum} & 100.0 & 63.8 & 88.9 & 0.0 & 42.1\\
Auror~\cite{shen16Auror} & 80.0 & 68.4 & 0.0 & 100.0 & \multicolumn{1}{c}{-}\\\hline
\textit{\ourname} & 0.0 & 62.0 & 100.0 & 100.0 & 100.0\\\hline

\end{tcolorbox}}
\captionsetup{type=table}
\captionof{table}{Experiment results of \ourname and six \sota defenses (three IR-based and three DF-based approaches) for the \cifar datasets in terms of Backdoor Accuracy ($BA$), Main Task Accuracy ($MA$), True-Positive-Rate (TPR), True-Negative-Rate (TNR), and Precision (PRC) for one FL round $t$, all values in percentage.}
\label{tab:successattacks}
\end{center}
\end{tablefloat}

\section{Discussion}
\label{sec:discussion}
\noindent 
In the earlier sections, we introduced \ourname, a novel defense against backdoors in FL that is fully compatible with secure aggregation techniques. In the following, we will discuss its security, parameters, as well as its limitations.

\subsection{Parameterization}
\label{sec:discussion:parameters}

 \noindent In our significance-based algorithm (cf. \hyperref[alg:significance]{\alg\ref{alg:significance}}), we introduce thresholds in the form of significance levels, that function as parameters for \ourname. The p-value of the probabilistic tests is set to 0.01 and the outlier thresholds are dynamic values based on the observed data. Thus, we purely rely on probabilistic thresholds and do not include empirically determined limits, that are dataset-dependent. We demonstrated that those parameters paired with \hyperref[alg:stackedclustering]{\alg\ref{alg:stackedclustering}} are robust against (un)targeted poisoning attacks. Naturally, higher p-values might stop the iterative validation although anomalous values are still present, hence increasing the probability of FNs, and, conversely, lower p-values make it more sensitive toward outliers leading to more rejected benign models and an increased \fpr.

\subsection{Necessity of TEEs}
\label{sec:discussion-tee}
\noindent \ourname requires the availability of a TEE on the server side and at clients. In the considered cross-silo scenario, where multiple larger computation centers collaborate on training a DNN, the availability of TEEs is reasonable to assume. While not all devices currently have TEEs, many of today's mobile devices possess TEEs (e.g., ARM-TrustZone). Although their deployment is restricted to vendors, this might change in the future. 
In scenarios with TEE-less clients it would be possible to, e.g., select only a subset of all clients for the feedback loop. Notably, in this case it is no longer guaranteed that the majority of the selected clients will be benign. In comparison, if all clients are used, this is guaranteed by the underlying threat model \hyperref[sec:problem-advmodel]{\sect\ref{sec:problem-advmodel}}. 
Our analysis provided in \app\ref{app:validationmajority} 
shows the probability of selecting a malicious majority in such a scenario. We note that for smaller PMRs it is negligible even if only 50 out of 1000 clients are selected for validation.  Therefore, in scenarios with TEE-less clients, stronger assumptions about the maximal PMR are necessary.

\subsection{Computational Overhead}
\label{sec:discussion-overhead}
\noindent\textbf{Overhead of Computations.} Validating the individual local models introduces an additional performance overhead. In \sect\ref{sec:eval-perf}, we evaluated this overhead and measured the runtimes of the individual phases of \ourname. Although the total runtime of 25.5s in average seems to be acceptable given much longer time needed for training a model, this overhead can still be further reduced. The major part of the overhead is created by the prediction of the DLOs. 
One strategy to improve the performance of \ourname would be utilizing ML accelerators, that include TEEs~\cite{nvidiaH100} or TEEs, that expand their security guarantees to accelerators~\cite{volos2018graviton, hetee}.
Also, the clients could decide to use only a representative subset of their local dataset for the validation. Other possible strategies to further optimize the runtime performance include parallelizing the calculations for the different models in each step, e.g., calculating the distances between the local models' and global model's DLO for different local models in parallel. Further improvements could be achieved by using a more performance-oriented language than Python. However, as the focus of this paper is on the design of a defense against poisoning attacks in FL, we consider those optimizations to be out of the scope. Therefore, it is left to future work to optimize the runtime performance of \ourname.

\noindent \textbf{Memory Overhead} Regarding memory, \ourname needs to hold the parameters of one DNN, the predicted DLOs of two DNNs, the DLOs for the global model, and for the one local model at the same time. In our setup for \cifar these aggregate to \numprint{94218} float-numbers per sample. After calculating the local model's DLOs, the distance to the global model DLOs can be determined and only these distances for each layer, i.e., a single number for each layer, need to be stored when continuing to process the next model. Depending on the available system resources, models that were not processed so far can be either stored (encrypted) on the file system, or the server sends the models sequentially to the clients. Hence, memory might be a limitation of \ourname in some TEEs. Nevertheless, newer architectures, such as SGXv2~\cite{li2022memory}, also provide large amounts of memory within an enclave.

\noindent\textbf{Overhead of Client-Side Validation.} The other aspect that causes the overhead is the distribution of the local updates to other clients. While the effort is negligible in the considered cross-silo scenario of a few collaborating computing centers, this overhead might become more relevant when applying \ourname to other scenarios with large numbers of participants. In \hyperref[sec:discussion-tee]{\sect\ref{sec:discussion-tee}}, we discussed the option to consider a subset of all clients for validation to the cost of stronger security assumptions required. Analogously, this can also be used to reduce the computation overhead. 

\section{Related Work}
\label{sec:sota}
\noindent In the recent past, a large number of backdoor attacks and defenses have been proposed. In the following, we discuss the approaches that are most relevant to this work and categorize them into the following types: DF approaches that aim to detect backdoored models (\hyperref[sec:sota-filtering]{\sect\ref{sec:sota-filtering}}) and IR approaches that mitigate the backdoors without identifying the poisoned models~(\hyperref[sec:sota-mitigation]{\sect\ref{sec:sota-mitigation}}). Afterwards, we will investigate privacy attacks and defenses~(\sect\ref{sec:sota-privacy}).

\subsection{Filtering Approaches}
\label{sec:sota-filtering}
\noindent Auror~\cite{shen16Auror} clusters selected parameters of the model updates on the server using k-means. In comparison, \ourname considers the outputs of all neurons. Additionally, Auror is vulnerable to multi-backdoor attacks where different clients inject different backdoors~\cite{nguyen22Flame}, like in our combined attack scenarios. The \mbox{significance-test-based} algorithm of \ourname handles such attacks by iteratively pruning the backdoored models.

FoolsGold~\cite{fung2020FoolsGold} assumes all clients to be \mbox{\nonIid}. For its analysis, it sums up all model updates that each client submitted to create a client-specific update history, before using the Cosine to compare the update histories of different clients. However, it can be circumvented by adaptive attacks~\cite{bagdasaryan} and fails to handle \iid scenarios. Further, the update history allows an adversary to gain trust by behaving benign for several rounds before performing its attack. Flame~\cite{nguyen22Flame} combines an \mbox{outlier-detection-based} approach with clipping and noising. However, the noising reduces the performance of the model, while the outlier detection fails in \nonIid scenarios. In comparison, \ourname does not affect the performance of the models and handles \iid and advanced \nonIid scenarios even with disjoint data. 

DeepSight~\cite{rieger2022deepsight} uses different techniques to extract fingerprints of the training data from the models' parameters and predictions to distinguish benign and backdoored models. Its classification relies on the assumption that poisoned models were trained on fewer labels as benign models. However, e.g., if the benign clients are trained only on a single label (cf. experiments for q=0.0 in \sect\ref{sec:eval}), this assumption does not hold, while \ourname can also in this corner case effectively distinguish benign and backdoored updates.

Zhao \etal~\cite{zhao2020shielding} is the closest to our work, as it also inspects the local models on the client side. The detection mechanism is based on the local models performance (MA) on the clients' local data, which means the output of the last layer only. To protect the privacy of the local models, Zhao \etal rely on applying differential privacy (DP) on the local models. In comparison, \ourname analyses the deep layer outputs and also detects sophisticated backdoors in models that do not affect the predictions for benign samples, therefore maintaining the MA, which is beneficial, since stealthy backdoor attacks do not affect MA (O2 in \hyperref[sec:problem-advmodel]{\sect\ref{sec:problem-advmodel}}). Further, performing the client-side analysis in TEEs guarantees the privacy of the local models but does not affect the analysis results. In comparison, applying DP, i.e., adding noise, changes the models and thus affects the results. Also, it is challenging to determine suitable parameters for DP, as too low parameters do not protect privacy while too strong parameters significantly affect the models' predictions and thus also the analysis results. Additionally, the clients have to report the number of samples for each label to the server to enable the server to choose suitable validation clients for each model~\cite{zhao2020shielding}. The filtering decision of Zhao~\etal is based on empirical thresholds instead of statistical tests as utilized in our approach. Hence, \ourname is independent of the dataset and model which results in a better real-world applicability.

FLARE utilizes a server-side dataset and applies KNN on a single layer’s plain outputs~\cite{wang2022flare}. However, assuming the presence of a server-side dataset is not practical~\cite{rieger2022deepsight}. In addition, focusing the analysis on a single layer allows the adversary to bypass the defense by fixing this single layer and hiding in others. Thus, \ourname is the first approach that uses all layers’ outputs to identify poisoned models without triggered samples.
\subsection{Mitigation Approaches}
\label{sec:sota-mitigation}
\noindent Other approaches try to mitigate the backdoor without identifying the poisoned models. Yin \etal~\cite{yin2018Median} proposed two approaches: One uses the parameters' median as a rule for aggregation, while the other one removes extreme parameter values and aggregates the remaining ones. Krum~\cite{blanchard17Krum} selects a single model as an aggregated model that minimizes the distance to a fraction of other models. However, the approaches of Yin \etal~\cite{yin2018Median} and Blanchard \etal~\cite{blanchard17Krum} do not work well in \nonIid scenarios, preventing \mbox{benign-but-outlier} models from being included in the aggregation. Naseri \etal~\cite{naseri2022local} propose using Differential Privacy (DP) for mitigating backdoors. However, besides the drawback that this strategy always reduces the model's performance, the DP level needs to be chosen manually. Here, a too high value makes the model unusable while a too low value is ineffective. In comparison, \ourname works without any \mbox{dataset-specific} parameters and relies on the significance level (p-value) of the statistical tests instead. Chen \etal look on models’ behavior during prediction and detects if input data triggering a backdoor are used~\cite{chen2018detecting}. For triggered samples, the predictions differ obviously, making detection straightforward. By leveraging \clientMetric, CrowdGuard’s validation algorithm allows detecting poisoned models without triggered data, even if the DLOs differ only marginally.
BaFFLe~\cite{andreina2020baffle} aggregates all local models and analyzes the final layer's output of the aggregated model via client feedback. Placing the backdoor detection after the aggregation allows the usage of secure-aggregation schemes. Therefore, BaFFLe relies for privacy protection on the security of the chosen secure-aggregation mechanism and the anonymity of the aggregated model. In contrast, \ourname inspects every layer of the individual local models on the client side. Therefore, BaFFLe's backdoor identification is based on a postulate that backdoors affect the predicted class for regular data which is not practical (see O2 in \hyperref[sec:problem-advmodel]{\sect\ref{sec:problem-advmodel}}). Further, BaFFLe cannot detect attacks in early rounds but needs multiple benign rounds for building a benign history, requires several empirically determined thresholds, and discards the whole training round if an attack is detected. In comparison, \ourname is round independent, uses only statistical thresholds based on probabilities that are independent of dataset and model, and allows training a model even in the presence of adversaries by filtering the poisoned updates.

In addition, all IR approaches have the disadvantage that the malicious clients cannot be identified and, hence, permanently excluded, but will permanently try to inject the backdoor, requiring the respective defense to always perfectly mitigate the poisoned models.  
 
\subsection{Privacy Attacks and Defenses}
\label{sec:sota-privacy}
\noindent \textbf{Privacy Attacks:} There are several attacks against ML models that are capable of leaking private information such as membership inference attacks~\cite{hayes2019logan,shokri2017membership}, property inference attacks~\cite{ganju2018property}, and label inference attacks~\cite{labelInference}. Additionally, inference methods that reconstruct the whole input have been developed~\cite{salem2019updates} not only for centralized ML processes but also in the area of FL~\cite{inferenceOnFL}. The TEE-based architecture of \ourname prevents inference attacks on the local models before the aggregation anonymizes the individual clients' contributions, while the aggregation of many models impedes inference attacks on the aggregated model~\cite{nasr2019comprehensive}. Thus, not only attacks are detected, but the real-world applicability of FL is pushed.

\noindent\textbf{Secure Aggregation Techniques:} Various approaches have been proposed to prevent \mbox{honest-but-curious}~\cite{fereidooni2021safelearn} or fully malicious servers~\cite{bonawitz,mo2021ppfl,hashemi2021byzantine} from accessing the local model updates. For example, Bonawitz \etal~\cite{bonawitz} use a secret-sharing protocol to allow the clients the calculation of noise that will cancel out during aggregation. However, this approach is not compatible with state-of-the-art backdoor defenses. Fereidooni \etal~\cite{fereidooni2021safelearn} use secure \mbox{multi-party} computation. However, these approaches create significant overhead for the clients and server. In the past, different approaches have been proposed using TEEs for the aggregation step. In PPFL~\cite{mo2021ppfl}, the whole FL process (training and aggregation) is performed inside a TEE. Hashemi \etal~\cite{hashemi2021byzantine} implemented Krum~\cite{blanchard17Krum} on SGX. In comparison to both, cryptograpy-based and TEE-based secure aggregation, \ourname not only implements secure aggregation inside a TEE. Instead, \ourname also provides an architecture to securely leverage clients' data for backdoor detection, without taking any privacy risk for local models or datasets.

\section{Conclusion}
\label{sec:con}
\noindent Privacy of sensitive data and defenses against poisoning attacks are central security considerations when it comes to Federated Learning (FL). To satisfy these needs, we propose \ourname, a model filtering defense against targeted poisoning attacks that introduces a client feedback loop leveraging the clients' local data for model assessment. In contrast to existing approaches, \ourname does not only rely on the vector metrics or models' accuracies but analyzes changes in the behavior of the deep layers' neurons to identify backdoor behavior. This enables \ourname to identify poisoned models independent of the clients' data distribution or the attack strategy.

\ourname has three core components:  1) A novel TEE-based architecture that allows using clients' data for the model validation without creating new privacy-attack vectors. 2) A \mbox{significance-based}  backdoor detection algorithm that executes statistical tests operating on \clientMetric, a novel metric based on the deep layer outputs of local models allowing to identify adversarial models. 3) A stacked clustering scheme, which compensates rogue votes of adversarial clients during the feedback loop. Thereby, our proposed architecture preserves the privacy, integrity, and confidentiality of local models and consequently client data by leveraging secure environments. 

We evaluate our approach in various FL settings and show the independence of those factors. Additionally, \ourname does not reduce the FL performance and is not circumventable by adaptive adversaries that are aware of the defense, making it applicable in \mbox{real-world} scenarios.


\section*{Acknowledgment}
This research received funding from Intel through the Private AI Collaborate Research Institute (\url{https://www.private-ai.org/}), as well as from the OpenS3 Lab and the Hessian Ministry of Interior and Sport as part of the F-LION project, following the funding guidelines for cyber security research.



%

\bibliographystyle{plain}
\bibliography{references}

\appendix
\section{Graphical Overview of \ourname}
\label{app:overview}
\begin{figure}[tb]
    \centering
    \includegraphics[width=.9\columnwidth]{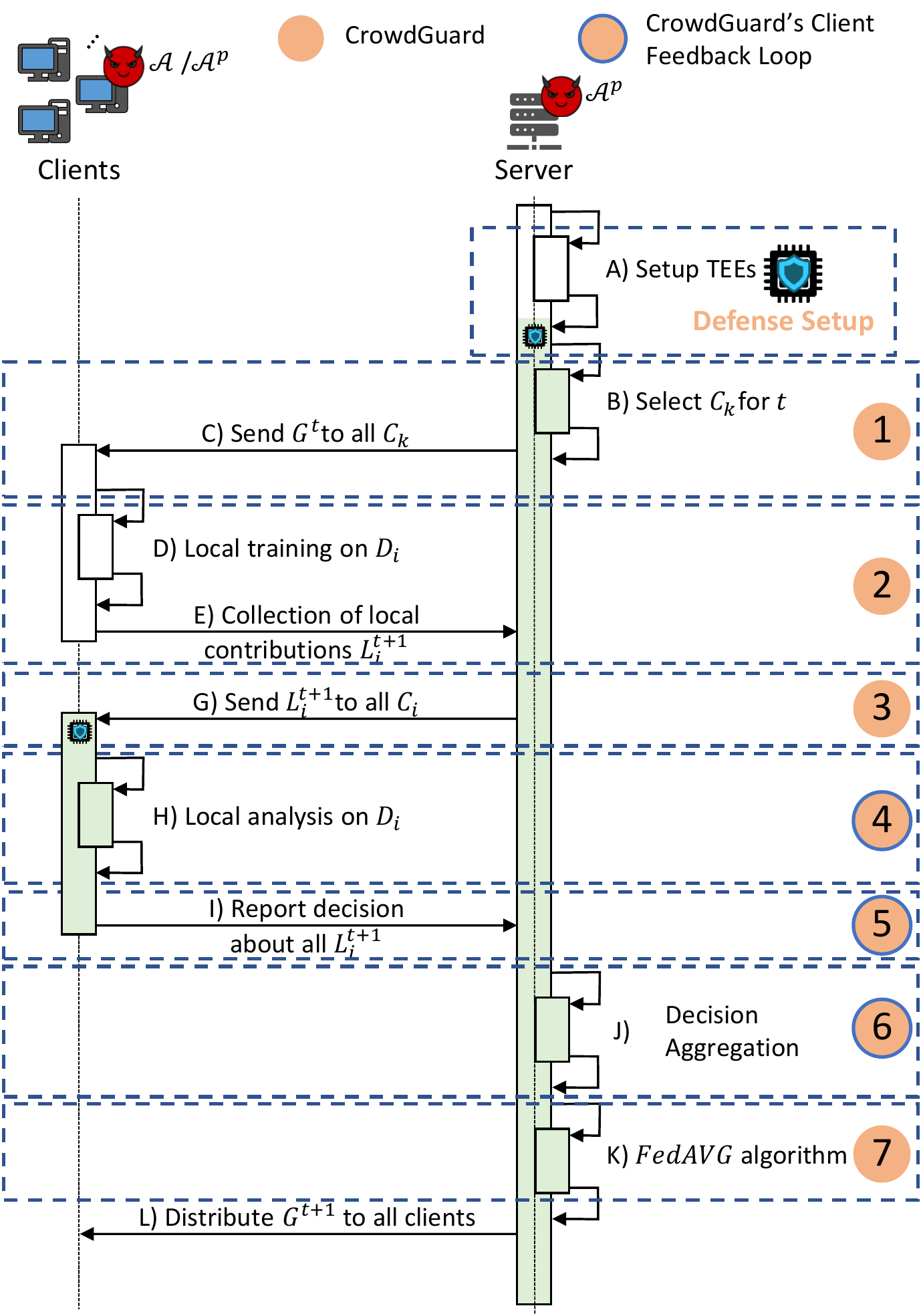}
    \caption{Execution sequence of \ourname. Green indicates the utilization of secure environments.}
    \label{fig:overallflow}
\end{figure}

\noindent\hyperref[fig:overallflow]{\fig\ref{fig:overallflow}} depicts a detailed overview of the FL setup with \ourname activated as a defense mechanism against targeted poisoning attacks.

\section{Significance Algorithm}
\label{app:algSignificance}
\noindent\hypertarget{alg:significance}{\alg\ref{alg:significance}} shows the algorithm that the individual clients execute during the validation to analyze the \clientMetric values and determine, whether there is still a significance for the presence of poisoned models in the calculated PCA values, such that \ourname needs to perform another pruning iteration.

It is also a valid proposition to consider leveraging multiple PC dimensions instead of solely relying on PC dimension one. However, our experiments have consistently demonstrated that the first PC dimension encompasses over 65\% (mostly exceeding 80\%) of the explained variances obtained through PCA. This measure signifies the significance and efficacy of this particular dimension in effectively segregating the data points. Incorporating the second dimension merely contributes to a marginal increase in the explained variances, which is why we opt to prioritize the utilization of the first dimension. This approach not only ensures computational efficiency but also maintains an intuitive framework.

An additional factor that influences our decision is the internal methods employed by \ourname. Specifically, the statistical tests implemented within \ourname are specifically designed to handle one-dimensional data. Employing multiple dimensions would necessitate the implementation of alternative algorithms, such as clustering methods. However, the statistical tests constitute the core of \ourname and are instrumental in yielding the positive effects associated with this defense mechanism.

\begin{algorithm}[thb]
	\caption{Significance Test on PC Values}
	\label{alg:significance} 

	\footnotesize
	{
	\begin{algorithmic}[1]

		\State \textbf{Input:}\\
		pc\_dim1\_values, \Comment{A list of values}
		
		\State \textbf{Output:}\\
		significant \Comment{Indicator if the values are contain abnormalities}
		
		\State \Comment{Generate distributions}
		\State median $\gets$ MEDIAN(pc\_dim1\_values)
		\State upper $\gets$ \{\}
		\State lower $\gets$ \{\}
		\For{value in pc\_dim1\_values}
		\State distribution\_value $\gets$ value $-$ median
		\If{value $>=$ 0}
		\State upper.append(distribution\_value)
		\Else
		\State lower.append(abs(distribution\_value))
		\EndIf
		\EndFor
		\State \Comment{Significance tests}
		\State mean\_significant $\gets$ T-TEST(upper, lower)
		\State var\_significant $\gets$ F-TEST(upper, lower)
		\State dist\_significant $\gets$ D-TEST(upper, lower)
		\State outlier\_quartil\_significant $\gets$ OUTLIER\_BOXPLOT(pc\_dim1\_values)
		\State outlier\_sigma\_significant $\gets$ OUTLIER\_3$\sigma$(pc\_dim1\_values)
		
		\State \Comment{Aggregate result}
		\State significant $\gets$ mean\_significant OR var\_significant OR dist\_significant OR outlier\_quartil\_significant OR outlier\_sigma\_significant
	\end{algorithmic}}

\end{algorithm}

\begin{figure*}[thb]
    \centering
    \subfloat[\label{fig:param:alpha}\centering $\alpha$]{{\includegraphics[height=2.6cm,trim={0 0.5cm 0 0}]{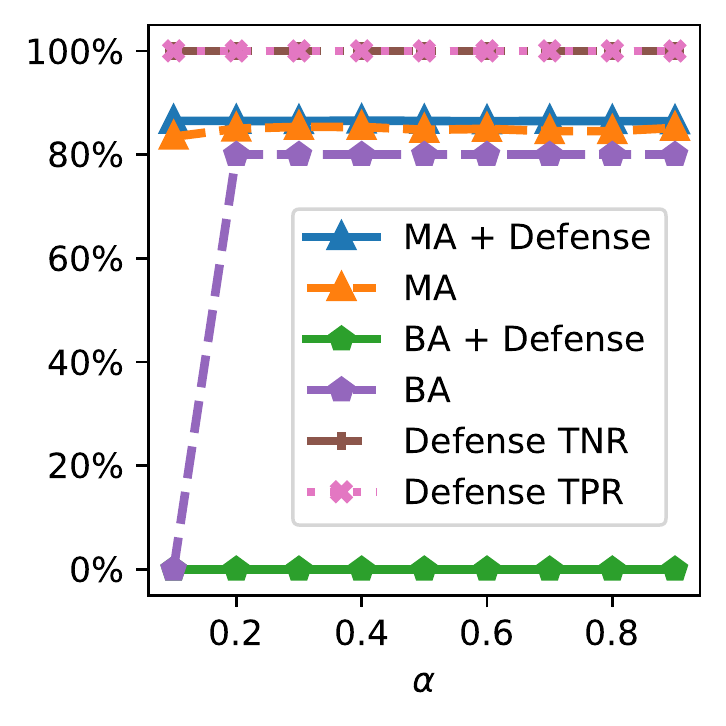} }}
    \qquad
    \subfloat[\label{fig:param:pdr}\centering PDR]{{\includegraphics[height=2.6cm,trim={0 0.5cm 0 0}]{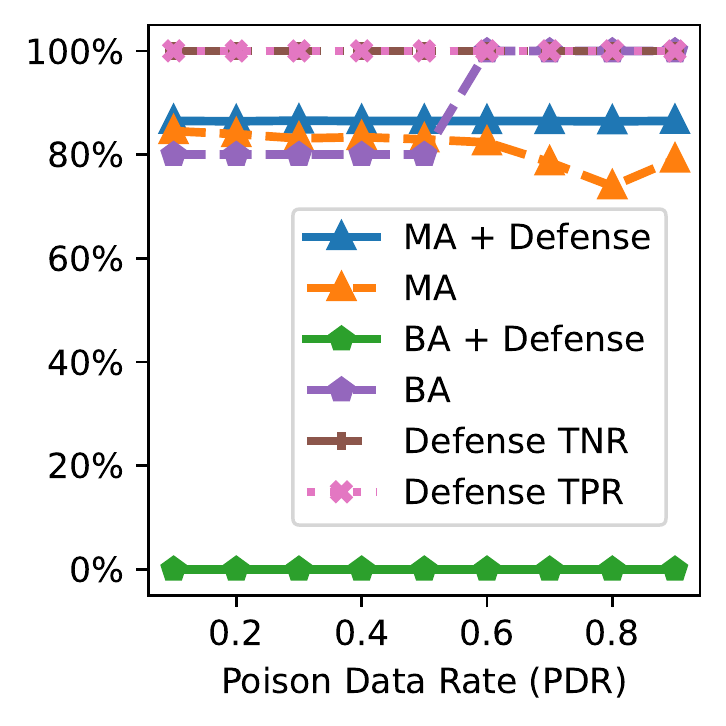} }}
    \qquad
    \subfloat[\label{fig:param:iid1}\centering \mbox{1-class} \nonIid]{{\includegraphics[height=2.6cm,trim={0 0.5cm 0 0}]{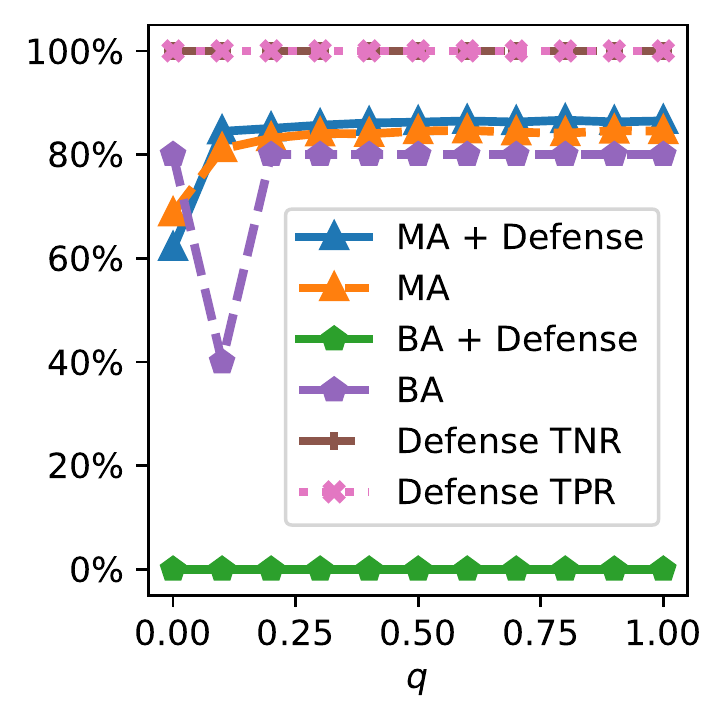} }}
    \qquad
    \subfloat[\label{fig:param:iid2}\centering \mbox{2-class} \nonIid]{{\includegraphics[height=2.6cm,trim={0 0.5cm 0 0}]{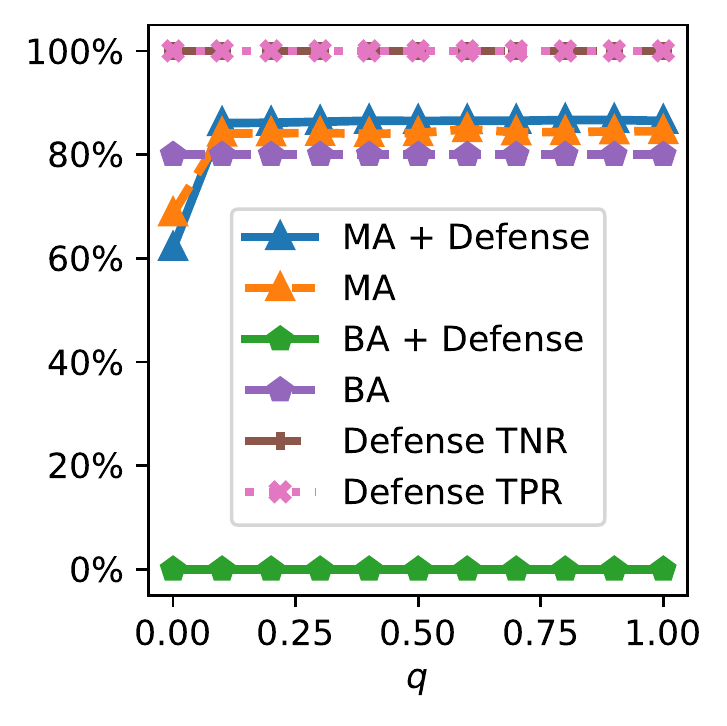} }}
    \qquad
    \subfloat[\label{fig:param:iid3}\centering Distributions]{{\includegraphics[height=2.6cm,trim={0 0.5cm 0 0}]{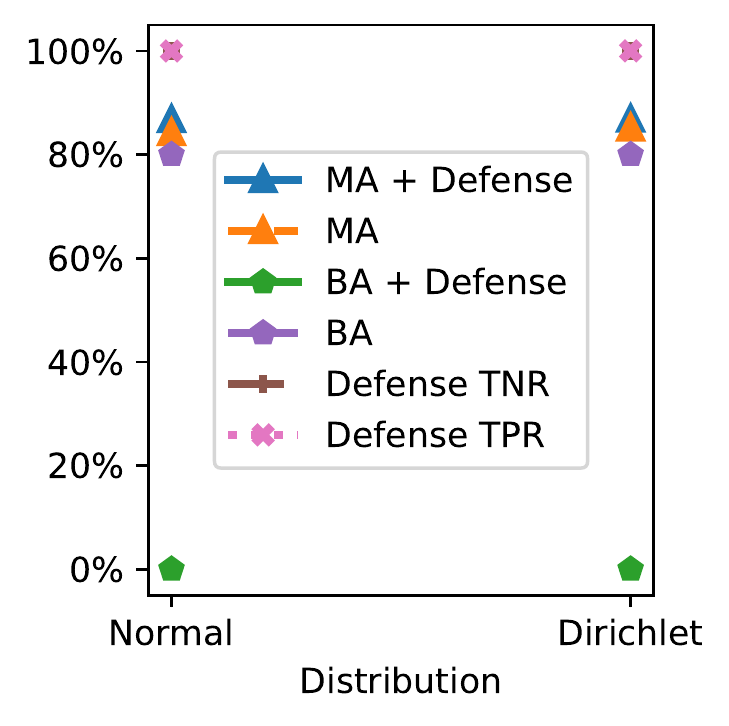} }}
    \caption{Influence of parameters.}
    \label{fig:param}
\end{figure*}

\section{Backdoor Types}
\label{app:backdoors}
\noindent Research knows about various kinds of triggers for different scenarios, i.e.~\cite{compositeTrigger,backdoorDataPoisoningNoise}, but we will only explain few of them, which are also used in our experiments:
\begin{itemize}
    \item \textbf{Pixel Backdoor}: This is a backdoor in the domain of image classification, where a pixel pattern is placed on the benign input image~\cite{bagdasaryan,badnets,trojanTriggerTargeted}, as visualized in \hyperref[fig:triggers:cartrigger]{\fig\ref{fig:triggers:cartrigger}} and the label is changed to the desired $T_\adversary$. In another injection strategy called \textit{Distributed Backdoor}, this trigger is distributed between multiple adversarial clients. Each client incorporates a fraction of the pattern into their local model. The final trigger is a combination of all the fractions~\cite{dbabackdoor}. 
    \item \textbf{Label Swap}: All samples of one label are swapped to $T_\adversary$. To create a poisoned dataset $\mathcal{D}^{\adversary}_i$, only changes regarding the label mapping are mandatory in $\mathcal{D}_i$.
    \item \textbf{Semantic Backdoor}: In this case, the input data contain a specific characteristic within the benign image, that should trigger a swap to $T_\adversary$. Examples regarding the CIFAR-10~\cite{krizhevsky2009learning} dataset are the mapping of cars in front of a striped background (cf. \hyperref[fig:triggers:cartrigger]{\fig\ref{fig:triggers:carmalicious}}) to $T_\adversary$, but leave all other car samples like \hyperref[fig:triggers:cartrigger]{\fig\ref{fig:triggers:car}} in its benign states~\cite{bagdasaryan}.
\end{itemize}

\section{Side-Channel Attacks on TEEs}
\label{app:sidechannels}
\noindent In the past, different side-channel attacks have been proposed that extract data from TEEs~\cite{sidechannelsgxusenix,huang2021aion,wang2018interface}. Different targets for such attacks in \ourname are possible. The direct target would be using side-channel attacks to first extract the local models of other clients from an enclave and then perform a model inference attack~\cite{ganju2018property,hayes2019logan,liu2022threats,nasr2019comprehensive,pyrgelis2017knock,salem2019updates,shokri2017membership,wang2019arxivEavesdrop,labelInference} on the extracted models. However, existing inference attacks have a low bandwidth of less than 100 bytes/s~\cite{van2021cacheout}, which is negligible compared to the size of a DNN model. 

Another option would be using such side-channel attacks to extract the cryptographic keys from the enclave and use this key to fake a TEE, thus to break the TEE completely. Also, an attacker could try to extract the keys that are used for the encrypted communication, i.e., the TLS session key, use the extracted key to eavesdrop the local models during their transmission and then run a model inference attack against the eavesdropped models. However, while attacks are proposed,  defenses against such attacks are also frequently developed. Examples against such side-channel attacks include the usage of constant-time encryption algorithms~\cite{barthe2020formal,daniel2020binsec} or techniques that randomized the data locations inside the memory~\cite{brasser2019dr,sang2022pridwen}. Thus, we consider such attacks to be out of the scope of this paper and assume TEEs to be trusted.

\section{Violation of the Benign Majority Assumption}
\label{app:validationmajority}
\begin{figure}[bt]
    \centering
    \includegraphics[width=0.9\columnwidth]{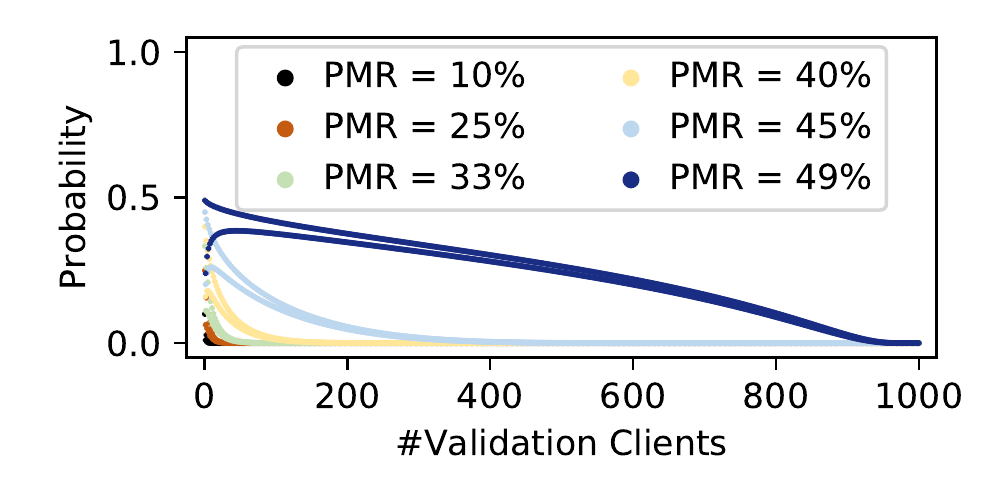}
    \caption{Probability for more than 50\% of adversaries being selected as validators out of 1000 clients for different PMRs.}
    \label{fig:percentage}
\end{figure}

\noindent The probability of violating the majority assumption, when a subset of all clients is randomly selected for validation, follows a hypergeometric distribution~\cite{rice2006mathematical}. 
 \hyperref[fig:percentage]{\fig\ref{fig:percentage}} shows  the probability that more than $50\%$ of the selected validation clients are malicious for different Poisoned Model Rates (PMRs) and different numbers of validation clients (for overall 1000 clients). As it can be seen, the probability for small PMR values becomes negligible already for less than 50 validation clients. 

\section{Experiment Parameter Setup}
\label{app:experimentparams}
\noindent In \hyperref[tab:defaultconf]{\tab\ref{tab:defaultconf}}, the default parameter configurations for our experiments are depicted.
\begin{tablefloat}[tb]
\begin{center}

\scaleTable{
\begin{tcolorbox}[tab3,tabularx={Z | Z},title=Default Configurations,boxrule=0.75pt]
Parameter & Default Value\\\hline\hline
 Dataset & CIFAR-10\\ 
 \hline
 Clients $n$ & $|C_k|$ = $|C_i|$ = 20\\ 
 \hline
 Epochs & 10\\ 
 \hline
 Samples per client & 2560\\ 
 \hline
 Batch Size & 64\\ 
 \hline
 Backdoor & Semantic Backdoor\\ 
 \hline
 IID rate $q$ & 0 \\ 
 \hline
 Poison Data Rate (PDR) & 0.1 \\
 \hline
 Starting round $t$ & 1000 \\
 \hline
 Adaptive adversary rate\footnotemark $\alpha$ & 0.7 \\
 \hline
 Poison Model Rate (PMR) & 0.45 ( = $\nicefrac{\mbox{9}}{\mbox{20}}$) \\
 \hline
 Benign Learning Rate & 0.01 \\
 \hline
 Malicious Learning Rate & 0.01\\
\end{tcolorbox}}
\captionsetup{type=table}
\captionof{table}{Listing of the default FL setup configurations.}
\label{tab:defaultconf}
\end{center}
\end{tablefloat}
\footnotetext{Adaptive adversary from Bagdasaryan \etal~\cite{bagdasaryan}.}

\section{Evaluation Metrics}
\label{app:metrics}

Besides the accuracy on the benign main task (Main Task Accuracy, MA) and the backdoor task (Backdoor Accuracy, BA), we also evaluate the filtering capabilities of \ourname by measuring the True Positive Rate (TPR) and True Negative Rate (TNR). For this purpose, we consider a benign model that is correctly recognized by \ourname to be a True Negative (TN), a poisoned model that is correctly identified as True Positive (TP) and analogously for False Positives (FP) and False Negatives (FN). The TPR is then defined as:
\begin{equation}
    \text{TPR} = \frac{\text{TP}}{\text{TP} + \text{FN}}
\end{equation}
The TNR is calculated as:
\begin{equation}
    \text{TNR} = \frac{\text{TN}}{\text{TN} + \text{FP}}
\end{equation}

\section{Evaluation of Vote Aggregation}
\label{sect:eval-clustering}

\begin{tablefloat}[t]
\begin{center}
\scaleTable{
\begin{tcolorbox}[tab5,tabularx={l||rr|rr|rr|rr|rr},boxrule=0.75pt]
	                \multirow{2}{*}{\textbf{Scenario}} & \multicolumn{2}{c|}{Majority} & \multicolumn{2}{c|}{K-Means} & \multicolumn{2}{c|}{Agglomerative} & \multicolumn{2}{c|}{DBSCAN} & \multicolumn{2}{c}{\ourname}\\
	                   &  TPR  &  TNR   &  TPR  &  TNR   &  TPR  &  TNR   &  TPR  &  TNR   &  TPR  &  TNR  \\\hline
	Default            & 100.0 &  100.0 & 100.0 &  100.0 & 100.0 &  100.0 & 100.0 &  100.0 & 100.0 &  100.0\\
	All Benign + 1 FN    &  88.9 &  100.0 &  88.9 &  100.0 &  88.9 &  100.0 & 100.0 &  100.0 & 100.0 &  100.0\\
	Default + 2 FP     & 100.0 &  100.0 & 100.0 &  100.0 & 100.0 &  100.0 &   0.0 &  100.0 & 100.0 &  100.0\\
	All Benign + 2 FN  &  77.8 &  100.0 &  77.8 &  100.0 &  77.8 &  100.0 &   0.0 &  100.0 & 100.0 &  100.0\\
	Malicious Split    & 100.0 &  100.0 &   0.0 &  100.0 &   0.0 &  100.0 & 100.0 &  100.0 & 100.0 &  100.0\\
\end{tcolorbox}}
\captionsetup{type=table}
\captionof{table}{Experimental result for the comparison of different aggregation rules for combining individual votes.}
\label{tab:voteaggregation}
\end{center}
\end{tablefloat}
\noindent As discussed in \sect\ref{sec:problem-advmodel}, we assume that the malicious clients can submit arbitrary votes by providing corresponding data to their enclave.

The stacked clustering ensures the integrity of the filtering, even if \adversary can manipulate the votes that are reported by the enclaves of the malicious clients (cf. \hyperref[alg:stackedclustering]{\alg\ref{alg:stackedclustering}}).
Tab.~\ref{tab:voteaggregation} shows the results of our ablation study for the vote aggregation of \ourname. We compare the stacked-clustering with different alternatives, in particular, majority voting, where a model is rejected if a majority of clients votes for its rejection, and K-Means, where the votes of the individual clients are clustered using K-Means and a model is accepted if at least one client in the majority cluster votes for acceptance. In addition, we evaluated the individual components of the stacked clustering (Agglomerative and DBSCAN) separately. We evaluate these aggregation rules for the votes that we observed for our default setting (Default). Further, we consider two scenarios, where all malicious clients vote for accepting all models but one/two benign clients do not detect one poisoned model (All Benign + 1 FN and All Benign + 2 FN). Further, we evaluate a scenario where all clients vote as observed in our default setting but two clients consider a single benign model to be poisoned (Default + 2 FP). In another setting, the adversary splits the clients into two groups: One group votes for accepting all models and one group for rejecting all models (Malicious Split). As \hyperref[tab:voteaggregation]{\tab\ref{tab:voteaggregation}} shows, only the stacked clustering of \ourname always detects all malicious models (TPR=100\%), while the other aggregation rules miss some of the models, or, in the case of K-Means, even accept all poisoned models.

\section{Ablation Study of \clientMetric}
To analyze the impact of using the relative distance in the \clientMetric metric, we conducted an additional experiment for an adapted version of \clientMetric, that utilizes absolute distances instead of relative versions. However, we observed that this adapted metric provides less insight into the models' behavior and it is more challenging to use these values to distinguish between benign and poisoned models.

\begin{figure}[tb]
    \centering
    \subfloat[\label{fig:abs}\centering Absolute]{{\includegraphics[height=2.8cm]{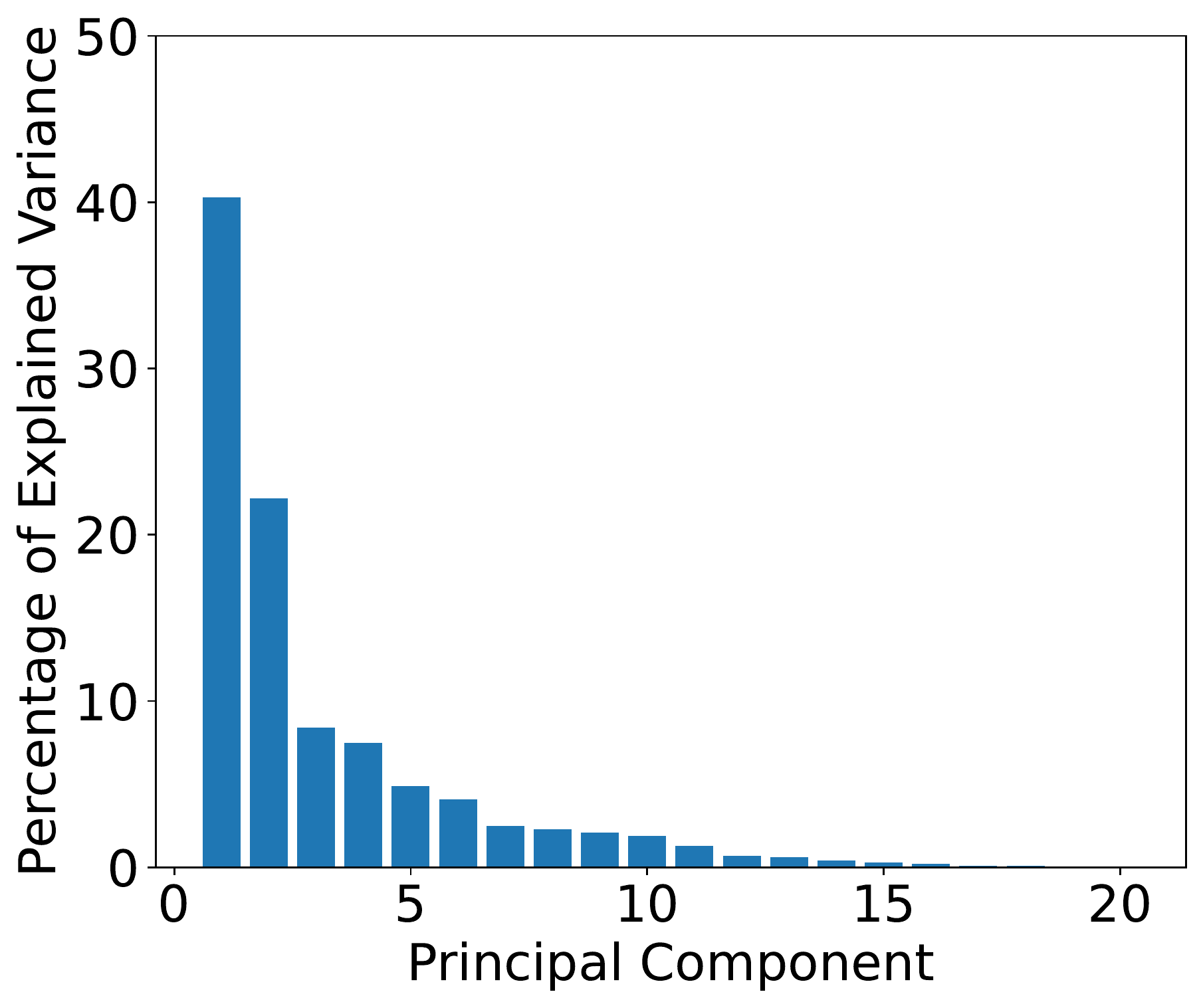} }}%
    \qquad
    \subfloat[\label{fig:rel}\centering Relative]{{\includegraphics[height=2.8cm]{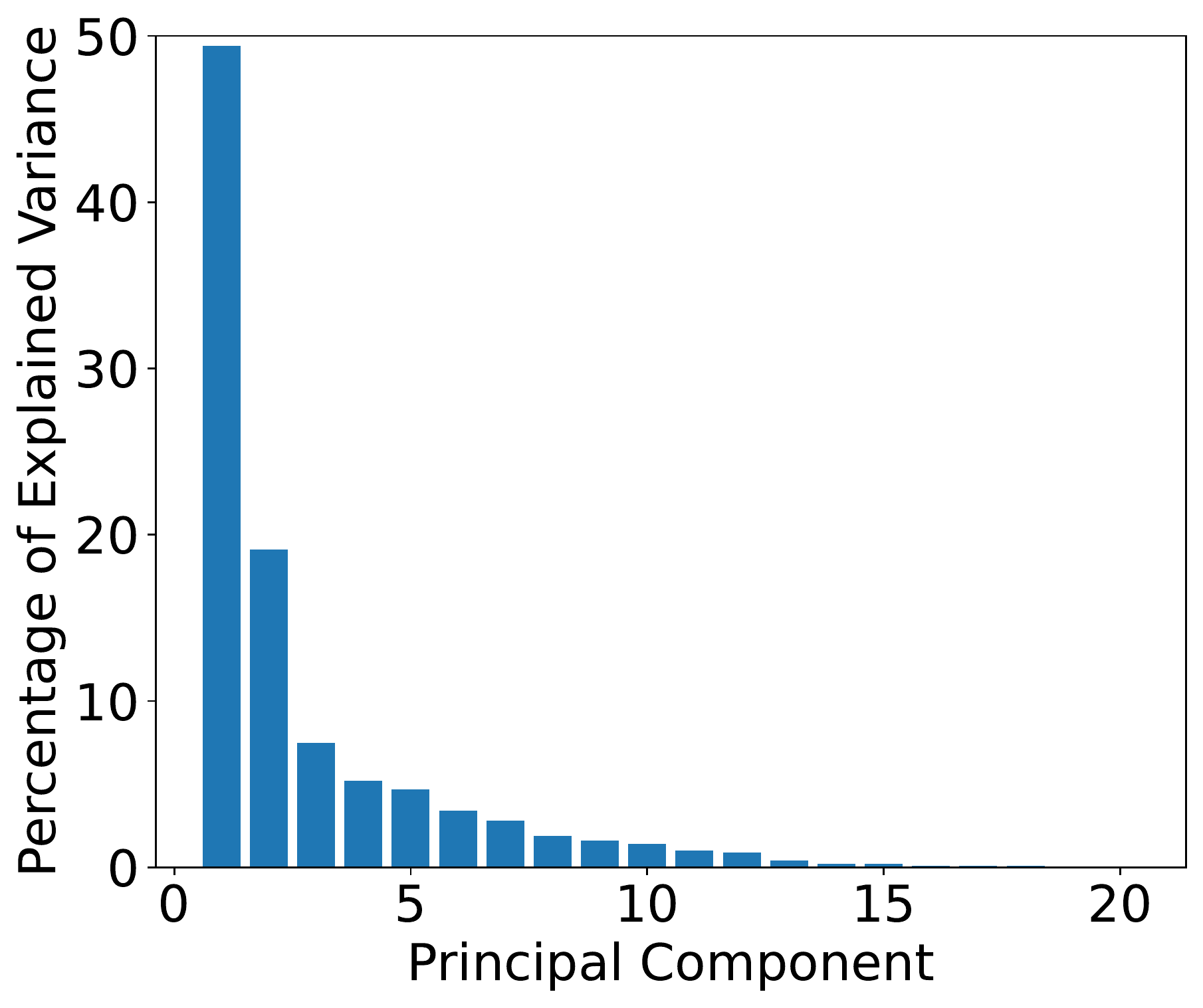} }}%
    \caption{Comparison of the expected variance of the PCA with absolute or relative distances used in the routine of \ourname. The figures are based on the Cosine distances in the first pruning round.}
    \label{fig:relabs}
\end{figure}

The reason for this is that utilizing relative distances allows us to consider the relative magnitude of updates. When a given parameter value is relatively small compared to another parameter that undergoes the same update value, the relative update becomes more relevant. After executing the PCA, this is also visible in the capability of the first Principal Component (PC) dimension to separate the data points, as measured by the explained variance. By leveraging the relative distances, we were able to enhance this capability by an average of 10\%, as can be seen in \hyperref[fig:relabs]{\fig\ref{fig:relabs}}. This improvement in separation capability resulted in better overall performance, particularly in edge cases. 

Thus, using relative distances proved to be more meaningful and beneficial for the HLBIM, allowing for improved detection and differentiation of poisoned models from benign values.

\section{Evaluation of Various Influence Factors}
\label{app:parameter-influence}
In \hyperref[fig:param]{\fig\ref{fig:param}}  
We depict graphs illustrating the influence of parameters in \hyperref[fig:param]{\fig\ref{fig:param}}. As it can be seen, the defense is independent of $\alpha$, the PDR, and the \nonIid scenario and achieves 100\% True-Positive-Rate (TPR) as well as True-Negative-Rate (TNR). The Main Task Accuracy (MA) is higher if the defense is activated, so we do not decrease the benign FL performance.

\end{document}